\documentclass[11pt,a4paper]{article}
\usepackage[a4paper,top=2.5cm,bottom=2.5cm,left=2.5cm,right=2.5cm]{geometry}
\usepackage{amsmath,amssymb}
\usepackage{newtxtext,newtxmath}
\usepackage{microtype}
\usepackage{graphicx}
\usepackage{booktabs}
\usepackage{flafter}
\usepackage{placeins}
\usepackage{caption}
\usepackage{titlesec}
\usepackage{enumitem}
\usepackage[colorlinks=true,linkcolor=blue,citecolor=blue,urlcolor=blue]{hyperref}

\titleformat{\section}{\large\bfseries\sffamily}{\thesection}{0.8em}{}
\titleformat{\subsection}{\normalsize\bfseries\sffamily}{\thesubsection}{0.8em}{}
\titleformat{\subsubsection}{\normalsize\bfseries\itshape\sffamily}{\thesubsubsection}{0.8em}{}
\titlespacing*{\section}{0pt}{2.2ex plus .5ex minus .2ex}{1.0ex}
\titlespacing*{\subsection}{0pt}{1.7ex plus .4ex minus .2ex}{0.7ex}
\titlespacing*{\subsubsection}{0pt}{1.3ex plus .3ex minus .2ex}{0.5ex}
\setlist{itemsep=1pt,topsep=3pt,parsep=1pt}
\newcommand{\tablelegend}[1]{\par\vspace{3pt}\begin{minipage}{\linewidth}\footnotesize #1\end{minipage}}
\makeatletter
\renewcommand{\@biblabel}[1]{#1.}
\makeatother
\let\origthebibliography\thebibliography
\renewcommand{\thebibliography}[1]{\small\origthebibliography{#1}}

\newcommand{\dd}{\mathrm{d}}
\newcommand{\bxi}{\boldsymbol{\xi}}
\newcommand{\bx}{\boldsymbol{x}}
\newcommand{\bu}{\boldsymbol{u}}
\newcommand{\bc}{\boldsymbol{c}}
\newcommand{\bn}{\boldsymbol{n}}
\newcommand{\bq}{\boldsymbol{q}}
\newcommand{\bj}{\boldsymbol{j}}
\newcommand{\bg}{\boldsymbol{g}}
\newcommand{\bpsi}{\boldsymbol{\psi}}
\newcommand{\bOm}{\boldsymbol{\Omega}}
\newcommand{\bJ}{\boldsymbol{J}}
\newcommand{\bU}{\mathbf U}
\newcommand{\bF}{\mathbf F}
\newcommand{\bI}{\mathbf I}
\newcommand{\avg}[1]{\left\langle #1\right\rangle}
\newcommand{\cS}{\mathcal S}
\newcommand{\cD}{\mathcal D}
\newcommand{\cT}{\mathcal T}
\newcommand{\sNS}{\boldsymbol\sigma_{\rm NS}}
\newcommand{\qNS}{\bq_{\rm NS}}
\newcommand{\stau}{\boldsymbol\sigma_{\tau}}
\newcommand{\qtau}{\bq_{\tau}}
\newlength{\panelh}
\AtBeginDocument{\setlength{\panelh}{0.36\textwidth}}

\begin{document}

\begin{center}
\parbox{0.84\linewidth}{\centering\Large\bfseries\sffamily A wave--particle decomposition framework for multiscale kinetic transport: continuous-spectrum equations and coupled wave--particle iteration\par}\par
\vspace{1.4em}
{\large Chang Liu$^{1,*}$ and Kun Xu$^{2,3,4}$\par}
\vspace{0.8em}
\begin{minipage}{0.92\linewidth}\small\centering
$^{1}$Institute of Applied Physics and Computational Mathematics, Beijing, China\\
$^{2}$Department of Mathematics, Hong Kong University of Science and Technology, Clear Water Bay, Kowloon, Hong Kong, China\\
$^{3}$Department of Mechanical and Aerospace Engineering, Hong Kong University of Science and Technology, Clear Water Bay, Kowloon, Hong Kong, China\\
$^{4}$Shenzhen Research Institute, Hong Kong University of Science and Technology, Shenzhen, China\\[3pt]
$^{*}$Corresponding author. E-mail addresses: \href{mailto:liuchang@iapcm.ac.cn}{liuchang@iapcm.ac.cn} (C. Liu), \href{mailto:makxu@ust.hk}{makxu@ust.hk} (K. Xu)
\end{minipage}
\end{center}
\vspace{0.8em}
\noindent\rule{\linewidth}{0.4pt}\par\smallskip
\noindent{\bfseries\sffamily Abstract}\par\smallskip
\noindent The wave--particle decomposition is presented as a framework for multiscale kinetic transport, comprising a theory and an algorithm. Along the characteristics of the kinetic equation, the equilibrium integral of the solution over a local kinetic horizon defines a wave, which depends only on the conservative variables, and its exact complement defines a particle. Both obey kinetic equations whose moments are extended Navier--Stokes equations, and the conservation laws together with the particle equation form a closed system that is equivalent to the kinetic equation for every horizon. Parametrized by the ratio of the horizon to the relaxation time, these systems form a continuous spectrum from kinetic theory to Navier--Stokes hydrodynamics that adapts to the flow from cell to cell. The algorithm is a coupled wave--particle iteration for steady flows. It alternates a macroscopic iteration for the conservative variables, with the particle frozen, and a kinetic iteration for the particle, with the wave fixed. Its fixed point is the discrete kinetic solution, it is asymptotic preserving, and its near-continuum convergence factor is bounded by the share of the transport carried by the particle, a share that decays exponentially with the horizon-to-relaxation ratio. The collision model enters only through a conservative remainder, and the equations and the iteration are written out for the Bhatnagar--Gross--Krook, Shakhov and ellipsoidal statistical models, the Boltzmann and Landau equations, neutron transport and radiative transfer. The decomposition is related to the unified gas-kinetic wave--particle method, the micro--macro decomposition, penalization and synthetic acceleration. Computations of hypersonic flows past a cylinder and around the three-dimensional X38 vehicle illustrate how the division of the distribution adapts to the flow and how strongly the iteration is accelerated.
\par\smallskip
\noindent\rule{\linewidth}{0.4pt}
\bigskip

\section{Introduction}\label{sec:intro}
Many flows of aerodynamic interest contain, within a single computational domain, regions where the mean free path is far smaller than the resolved length and regions where the two are comparable. Around a vehicle at high altitude, the shock layer and the boundary layer lie close to the continuum limit, while the expansion over the shoulder and the wake are rarefied. A plume expanding into vacuum or the flow through a microchannel presents the same contrast. A kinetic equation describes all of these regions, yet solving it directly is most expensive where the kinetic detail matters least. In near-continuum regions the collision scale must be resolved unless the scheme is asymptotic preserving, that is, unless on a coarse mesh it reduces to a consistent discretization of the hydrodynamic limit~[1, 2]. Steady problems add a second obstacle. The conventional iteration, which inverts the transport with the collision term lagged, needs a number of iterations that grows like the inverse square of the Knudsen number~[3].

Both difficulties have been approached from several directions. Asymptotic-preserving time discretizations treat the stiff relaxation implicitly and retain the fluid limit at the discrete level~[4, 5, 6, 7]. For a general collision operator, Filbet and Jin~[8] penalize the operator by a Bhatnagar--Gross--Krook (BGK) term, which is integrated implicitly, and integrate the remainder explicitly. Micro--macro decompositions~[9, 10, 11, 12] write the distribution as a Maxwellian plus a kinetic remainder and couple a macroscopic conservation law with a kinetic equation for that remainder, which opens a systematic route to asymptotic-preserving schemes with exact conservation of moments. Hybrid methods, finally, assign a kinetic description to some regions and a hydrodynamic one to others and pass between them through a local criterion~[13, 14].

A different line of work builds the numerical method on the integral solution of the kinetic equation. The unified gas-kinetic scheme (UGKS)~[15, 16, 17] and the discrete unified gas-kinetic scheme~[18] construct the numerical flux from this solution, which couples transport and collisions on the scale of the mesh and makes the scheme asymptotic preserving~[19]. The UGKS has since been extended to the full Boltzmann operator~[20], to radiative transfer~[21], to plasmas~[22] and to multiphase flows~[23]. Its wave--particle formulation, the unified gas-kinetic wave--particle (UGKWP) method~[24, 25, 26], represents the same integral solution within each time step by an analytically evolved wave and by stochastic particles, so that the number of kinetic degrees of freedom follows the local ratio of the time step to the collision time~[27]. For steady problems, the slow convergence of the conventional iteration is overcome by synthetic acceleration, in which each transport sweep is corrected by the solution of a low-order equation. The diffusion synthetic acceleration of neutron transport~[28, 29, 3] is the classical example. In the implicit UGKS~[30, 31, 32], implicit macroscopic equations predict the conservative variables before the distribution is updated, and the general synthetic iterative scheme~[33, 34] solves synthetic moment equations, closed by the kinetic solution, between kinetic iterations.

The wave--particle decomposition~[35, 36, 37] takes another route. Instead of coupling the kinetic equation to a separate fluid model or dividing the domain, it reformulates the kinetic equation itself. The starting point is again the integral solution along characteristics. To every point we attach a time interval $\ell$, the local kinetic horizon, and split the integral solution over this horizon into two contributions, the Maxwellians emitted by collisions within the horizon and the distribution that enters the horizon from upstream. The first contribution, the wave $W$, depends on the solution only through the conservative variables $\bU$ along the characteristic. The second, the particle $P$, is defined as the exact complement, so that the identity $f=W+P$ involves no approximation. When the horizon is long compared with the relaxation time, almost the whole distribution is wave and the description is essentially macroscopic. When it is short, the wave disappears and the particle is the distribution function itself.

The collision operator enters this construction only through an identity that writes it as a BGK relaxation towards the local Maxwellian plus a conservative remainder $R$. The wave is therefore the same for every collision model, and the model affects the particle alone. The remainder vanishes for the BGK model, corrects the relaxation target explicitly for the Shakhov and the ellipsoidal statistical BGK (ES-BGK) models, and contains the full collision operator for the Boltzmann and Landau equations. Linear transport has the same structure, with the isotropic source function in place of the Maxwellian and the diffusion equation in place of the Navier--Stokes equations, and neutron transport and radiative transfer are included for this reason. Because the collision model enters the equations and the algorithm alike through a few well-defined parts, the decomposition is better described as a framework than as a scheme for one particular equation.

The theory is built from three sets of equations. The wave and the particle obey kinetic equations that exchange the Maxwellian leaving the horizon. Their velocity moments are extended Navier--Stokes equations, in which each component has its own density, pressure and diffusion flux and the two components exchange mass, momentum and energy. Adding the moment equations gives the conservation law for $\bU$, whose flux is the sum of the wave flux, a functional of $\bU$, and the particle flux, a moment of $P$. Together with the particle equation, this law forms a closed system for $(\bU,P)$, the wave--particle multiscale equations, which is equivalent to the kinetic equation for every horizon. Near equilibrium its fluxes are known in closed form. In terms of the ratio $\eta=\ell/\tau$ of the horizon to the relaxation time, the wave carries the fraction $1-e^{-\eta}$ of the Euler flux and the fraction $1-(1+\eta)e^{-\eta}$ of the viscous flux generated by the relaxation term, and the particle carries the rest. As $\eta$ grows from zero to infinity, the system passes continuously from the kinetic equation to the Navier--Stokes equations, in which the particle merely corrects the transport coefficients. Since $\eta$ is a local quantity, this continuous spectrum of equivalent systems adapts to the flow from cell to cell.

The algorithm solves the steady multiscale equations by a coupled wave--particle (WP) iteration. A macroscopic iteration, the W iteration, advances the conservation law for $\bU$ with the particle frozen. It evaluates no collision term, and its cost does not depend on the velocity grid. A kinetic iteration, the P iteration, advances the particle equation with the wave fixed and evaluates the collision term once. The two are linked by an endpoint traction, which passes the macroscopic prediction on to the particle, and by an acceptance step, which redefines $\bU$ as the moment of the updated distribution. Whatever the parameters of the iteration, the converged state satisfies the discrete kinetic equation. The scheme is asymptotic preserving, and in the near-continuum regime its convergence factor is bounded by the share $(1+\eta)e^{-\eta}$ of the transport carried by the particle, which decays exponentially with $\eta$.

The decomposition was introduced in~[35] for kinetic relaxation models and realized there by an explicit scheme with a gas-kinetic wave flux and discrete-ordinate or Monte Carlo particles. The coupled WP iteration was then developed for steady flows of the Shakhov model in~[36] and of the full Boltzmann equation in~[37]. Those papers concentrate on the schemes and their numerical assessment, while the present paper concentrates on the framework itself. It collects the equations and the algorithm in a form that does not depend on the collision model, so that a further model enters through a small number of clearly identified parts, states the properties that follow from this structure, and places the decomposition among related methods.

The paper is organized as follows. Section~2 develops the theory, from the general kinetic equation and the wave operator through the microscopic and macroscopic equations of the two components to the wave--particle multiscale equations and their continuous spectrum, and closes with the equations of seven specific models. Section~3 presents the coupled WP iteration in the same order, from its architecture to the schemes for the seven models. Section~4 states the asymptotic-preserving and acceleration properties and relates the decomposition to the UGKWP method, the micro--macro decomposition, penalization and synthetic acceleration. Section~5 illustrates the framework with hypersonic flows past a cylinder and around the three-dimensional X38 vehicle, from the transitional to the continuum regime, and with a lid-driven cavity in the continuum regime. Section~6 concludes.

\section{Theoretical equations}\label{sec:method}

\subsection{General kinetic equation}
We consider a monatomic gas described by the distribution function $f(\bx,\bxi,t)$ of position $\bx$, molecular velocity $\bxi\in\mathbb R^3$ and time $t$, which evolves according to
\begin{equation}\label{eq:kinetic}
 \partial_tf+\bxi\cdot\nabla_{\bx}f=\frac{1}{\varepsilon}\,C(f),
\end{equation}
where $\varepsilon$ is the Knudsen number and $C$ the collision term. All quantities are nondimensional, and the temperature is scaled so that the pressure is $p=\rho T$. Velocity moments are denoted by $\avg{\cdot}=\int_{\mathbb R^3}\cdot\,\dd\bxi$, and $\bpsi=(1,\bxi^T,|\bxi|^2/2)^T$ are the collision invariants. The conservative variables are
\begin{equation}
 \bU=\avg{\bpsi f}=(\rho,\rho\bu^T,\rho E)^T,
\end{equation}
with the specific total energy $E=|\bu|^2/2+3T/2$, and in terms of the peculiar velocity $\bc=\bxi-\bu$ the local Maxwellian reads
\begin{equation}
 M=\frac{\rho}{(2\pi T)^{3/2}}\exp\Big(-\frac{|\bc|^2}{2T}\Big).
\end{equation}
Only three properties of the collision term are needed. It conserves mass, momentum and energy, so that $\avg{\bpsi\,C(f)}=\boldsymbol0$. It vanishes on the local Maxwellian, $C(M)=0$. Finally, in smooth flow it admits the Chapman--Enskog expansion $f=M+\varepsilon f^{(1)}+\mathcal O(\varepsilon^2)$, whose moments give the Navier--Stokes equations of the model with its viscosity $\mu$, heat conductivity $\kappa$ and Prandtl number $\Pr=c_p\mu/\kappa$, $c_p=5/2$.

The decomposition rests on a simple rewriting of the collision term. Let $\tau(\bx,t)>0$ be a relaxation time that does not depend on $\bxi$. Adding and subtracting the BGK relaxation term~[38] turns (1) into
\begin{equation}\label{eq:split}
 \begin{aligned}
 &\partial_tf+\bxi\cdot\nabla_{\bx}f=\frac{M-f}{\tau}+R,\\
 &R=\frac{1}{\varepsilon}\,C(f)-\frac{M-f}{\tau},
 \end{aligned}
\end{equation}
which is an identity and involves no modelling. Because $M$ and $f$ share their conservative moments, the collision remainder $R$ is conservative,
\begin{equation}
 \avg{\bpsi R}=\frac1\varepsilon\avg{\bpsi\,C(f)}-\frac1\tau\big(\avg{\bpsi M}-\avg{\bpsi f}\big)=\boldsymbol0,
\end{equation}
and it vanishes for $f=M$. The collision model enters the theory through this remainder alone, which for the models considered below reads
\begin{equation}\label{eq:remainders}
 \begin{aligned}
 &\text{BGK model:} && R=0,\\
 &\text{Shakhov model:} && R=(S-M)/\tau,\\
 &\text{ES-BGK model:} && R=(G-M)/\tau,\\
 &\text{Boltzmann equation:} && R=Q(f,f)/\varepsilon-(M-f)/\tau,\\
 &\text{Landau equation:} && R=Q_L(f,f)/\varepsilon-(M-f)/\tau.
 \end{aligned}
\end{equation}
For a relaxation model, $\tau$ is the relaxation time of the model. For the Boltzmann and Landau equations it is a free parameter of the decomposition that leaves $f$ unchanged, and we make the choice $\tau=\mu/p$ below. Linear transport, in which a source function takes the place of the Maxwellian, is treated in Sections~2.7.6 and~2.7.7.

\subsection{Wave operator}
With the identity (4) in hand, the wave is built from its integral solution along characteristics. Fix a point $(\bx,\bxi,t)$ and measure the flight time $s\ge0$ backwards along the characteristic. A subscript $s$ denotes evaluation at the characteristic point, $a_s=a(\bx-\bxi s,\,t-s)$, and the cumulative collision frequency $\nu(s)=\int_0^s\tau_r^{-1}\,\dd r$ defines the collisionless factor $e^{-\nu(s)}$, the probability that a molecule travels for the time $s$ without colliding. Along the characteristic, (4) and the definition of $\nu$ give
\begin{equation}
 \begin{aligned}
 &\frac{\dd f_s}{\dd s}=-\frac{M_s-f_s}{\tau_s}-R_s,\\
 &\frac{\dd}{\dd s}\Big(e^{-\nu(s)}f_s\Big)=-e^{-\nu(s)}\Big(\frac{M_s}{\tau_s}+R_s\Big).
 \end{aligned}
\end{equation}
Let $\ell(\bx,t)\ge0$ be the local kinetic horizon, which is measured in time. Integrating the second relation over $0\le s\le\ell$ yields the integral solution
\begin{equation}\label{eq:integral}
 f=\int_0^{\ell}\frac{M_s}{\tau_s}e^{-\nu(s)}\,\dd s+e^{-\nu(\ell)}f_\ell+\int_0^{\ell}R_se^{-\nu(s)}\,\dd s .
\end{equation}
Its first term, the equilibrium integral over the horizon, defines the wave operator, and the particle is the exact complement,
\begin{equation}\label{eq:wave}
 \begin{aligned}
 &W=\int_0^{\ell}\frac{M_s}{\tau_s}e^{-\nu(s)}\,\dd s,\\
 &P=f-W=e^{-\nu(\ell)}f_\ell+\int_0^{\ell}R_se^{-\nu(s)}\,\dd s .
 \end{aligned}
\end{equation}
Since $M$ and $\tau$ are functions of $\bU$, the wave is a functional of $\bU$ on the backward characteristic segment of duration $\ell$, truncated at the initial time or at an inflow boundary, and it never becomes an independent unknown. The particle collects the upstream distribution that reaches $(\bx,t)$ without a collision, together with the accumulated remainder (Fig.~1). When $R\neq0$ it carries a signed contribution and need not be non-negative, since non-negativity is required of $f$ alone. The decomposition is exact for every horizon, and its character is governed by the horizon-to-relaxation ratio
\begin{equation}
 \eta=\frac{\ell}{\tau}.
\end{equation}
\begin{figure}[tbp]
\centering
\includegraphics[width=0.62\textwidth]{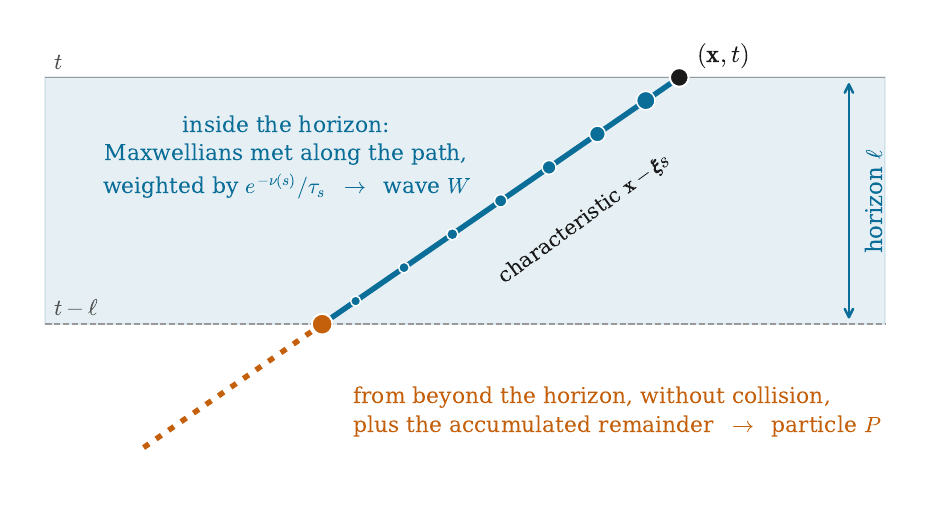}
\caption{The wave and the particle at a point $(\bx,t)$. Along the backward characteristic, the Maxwellians met within the horizon $\ell$, weighted by the collisionless factor, form the wave $W$; the distribution that enters the horizon from beyond it without a collision, together with the accumulated collision remainder, forms the particle $P$.}\label{fig:sketch}
\end{figure}

\subsection{Wave equations}
Having defined the wave, we ask how it evolves. At the microscopic level, its evolution follows from differentiating (9) along the characteristic. Let $\cD=\partial_t+\bxi\cdot\nabla_{\bx}$ act on $(\bx,t)$ at fixed $s$. Every field satisfies $\cD a_s=-\partial_sa_s$, and $\cD\nu(s)=1/\tau-1/\tau_s$. The integrand $K(s)=M_se^{-\nu(s)}/\tau_s$ of the wave, for which $K(0)=M/\tau$, therefore obeys
\begin{equation}
 \cD K(s)=-\partial_sK(s)-\frac{K(s)}{\tau}.
\end{equation}
and the Leibniz rule gives
\begin{equation}
 \cD W=K(\ell)\,\cD\ell+\int_0^\ell\cD K\,\dd s=K(\ell)\,\cD\ell-K(\ell)+K(0)-\frac{W}{\tau},
\end{equation}
which is the wave equation
\begin{equation}\label{eq:wave_equation}
 \begin{aligned}
 &\partial_tW+\bxi\cdot\nabla_{\bx}W=\frac{M-W}{\tau}-\cS_\ell,\\
 &\cS_\ell=\frac{M_\ell}{\tau_\ell}\,e^{-\nu(\ell)}\big(1-\partial_t\ell-\bxi\cdot\nabla_{\bx}\ell\big).
 \end{aligned}
\end{equation}
The wave relaxes towards the Maxwellian at the rate $1/\tau$ and loses the horizon exchange $\cS_\ell$, the rate at which Maxwellian contributions leave the horizon. The collision model does not appear in this equation at all.

At the macroscopic level, it is convenient to take the moments of both components $Z=W,P$ relative to the bulk velocity $\bu$ of $f$. The density, diffusion flux, total energy, pressure, stress and heat flux of a component are
\begin{equation}\label{eq:component_moments}
 \begin{aligned}
 &\rho_Z=\avg{Z},\qquad \bj_Z=\avg{\bc\,Z},\qquad (\rho E)_Z=\tfrac12\avg{|\bxi|^2Z},\qquad p_Z=\tfrac13\avg{|\bc|^2Z},\\
 &\boldsymbol\sigma_Z=\avg{\big(\bc\bc^T-\tfrac13|\bc|^2\bI\big)Z},\qquad \bq_Z=\tfrac12\avg{|\bc|^2\bc\,Z},
 \end{aligned}
\end{equation}
so that $\bU_Z=\avg{\bpsi Z}=(\rho_Z,(\rho_Z\bu+\bj_Z)^T,(\rho E)_Z)^T$, and its flux $\bF_Z=\avg{\bpsi\bxi^TZ}$ contains the generalized stress and heat flux
\begin{equation}\label{eq:generalized}
 \widetilde{\boldsymbol\sigma}_Z=\boldsymbol\sigma_Z+\bu\bj_Z^T+\bj_Z\bu^T,\qquad
 \widetilde\bq_Z=\bq_Z-(\bu\cdot\bj_Z)\,\bu-\tfrac12|\bu|^2\bj_Z .
\end{equation}
Since $\avg{\bpsi M}=\bU$, the moments of the right-hand side of (13) are
\begin{equation}\label{eq:exchange_source}
 \boldsymbol{\mathcal S}=(s_\rho,\boldsymbol s_m^T,s_E)^T=\frac{\bU_P}{\tau}-\avg{\bpsi\,\cS_\ell},
\end{equation}
in which relaxation converts particle into wave at the rate $1/\tau$ and the horizon exchange returns the part that leaves the horizon. The moments of (13) are therefore the macroscopic wave equations
\begin{equation}\label{eq:wave_macro}
 \left\{
 \begin{aligned}
 &\partial_t\rho_W+\nabla_{\bx}\cdot\big(\rho_W\bu+\bj_W\big)=s_\rho,\\
 &\partial_t\big(\rho_W\bu+\bj_W\big)+\nabla_{\bx}\cdot\big(\rho_W\bu\bu^T+p_W\bI+\widetilde{\boldsymbol\sigma}_W\big)=\boldsymbol s_m,\\
 &\partial_t(\rho E)_W+\nabla_{\bx}\cdot\big[\big((\rho E)_W+p_W\big)\bu+\widetilde{\boldsymbol\sigma}_W\bu+\widetilde\bq_W\big]=s_E .
 \end{aligned}
 \right.
\end{equation}
These have the form of extended Navier--Stokes equations. The wave possesses its own density, pressure and diffusion flux, its stress and heat flux are generalized by the diffusion flux, and it receives mass, momentum and energy from the particle.

\subsection{Particle equations}
At the microscopic level, subtracting (13) from (4) and using $f-W=P$ gives the particle equation
\begin{equation}\label{eq:particle_equation}
 \partial_tP+\bxi\cdot\nabla_{\bx}P=R-\frac{P}{\tau}+\cS_\ell=R+\frac{P_e-P}{\tau},\qquad P_e=\tau\cS_\ell .
\end{equation}
The particle carries the collision remainder and relaxes at the rate $1/\tau$ towards the endpoint distribution $P_e$, which for a constant horizon and a locally frozen relaxation time is $P_e=e^{-\eta}M_\ell$. This is the only equation in which the collision model appears. Since the wave and particle equations add up to (4), the full collision term is retained for every horizon. For $\ell=0$ the wave vanishes, $\cS_\ell=M/\tau$ and $P=f$, and (18) is the kinetic equation itself.

At the macroscopic level, the conservation property of the remainder removes the collision term from the moments of (18). With the quantities (14) and (15) taken for $Z=P$, these moments are the macroscopic particle equations
\begin{equation}\label{eq:particle_macro}
 \left\{
 \begin{aligned}
 &\partial_t\rho_P+\nabla_{\bx}\cdot\big(\rho_P\bu+\bj_P\big)=-s_\rho,\\
 &\partial_t\big(\rho_P\bu+\bj_P\big)+\nabla_{\bx}\cdot\big(\rho_P\bu\bu^T+p_P\bI+\widetilde{\boldsymbol\sigma}_P\big)=-\boldsymbol s_m,\\
 &\partial_t(\rho E)_P+\nabla_{\bx}\cdot\big[\big((\rho E)_P+p_P\big)\bu+\widetilde{\boldsymbol\sigma}_P\bu+\widetilde\bq_P\big]=-s_E,
 \end{aligned}
 \right.
\end{equation}
which differ from (17) only in the sign of the exchange. Both systems keep the same form for every collision model.

\subsection{Wave--particle multiscale equations}\label{sec:closed}
Because $f=W+P$, the component moments add up, $\rho_W+\rho_P=\rho$, $\bj_W+\bj_P=\boldsymbol0$, $p_W+p_P=p$ and $(\rho E)_W+(\rho E)_P=\rho E$. When (17) and (19) are added, the diffusion corrections and the exchange cancel, and what remains is the total conservation law $\partial_t\bU+\nabla_{\bx}\cdot(\bF_W+\bF_P)=\boldsymbol0$. The wave is determined by $\bU$, so that this law and the particle equation form a closed system,
\begin{equation}\label{eq:closed}
 \left\{
 \begin{aligned}
 &\partial_t\rho+\nabla_{\bx}\cdot(\rho\bu)=0,\\
 &\partial_t(\rho\bu)+\nabla_{\bx}\cdot\big(\rho\bu\bu^T+p\bI+\boldsymbol\sigma_W+\boldsymbol\sigma_P\big)=\boldsymbol0,\\
 &\partial_t(\rho E)+\nabla_{\bx}\cdot\big[(\rho E+p)\bu+(\boldsymbol\sigma_W+\boldsymbol\sigma_P)\bu+\bq_W+\bq_P\big]=0,\\
 &\partial_tP+\bxi\cdot\nabla_{\bx}P=R+\frac{P_e-P}{\tau},
 \end{aligned}
 \right.
\end{equation}
in which the collision remainder and the endpoint distribution are
\begin{equation}\label{eq:closure}
 \begin{aligned}
 &R=\frac{1}{\varepsilon}\,C(f)-\frac{M-f}{\tau},\\
 &P_e=\tau\cS_\ell=\frac{\tau}{\tau_\ell}\,M_\ell\,e^{-\nu(\ell)}\big(1-\partial_t\ell-\bxi\cdot\nabla_{\bx}\ell\big),
 \end{aligned}
\end{equation}
while $W$ is given by (9) with $M$ the Maxwellian of $\bU$, and $f=W+P$. These are the wave--particle multiscale equations, on which the method rests. Their unknowns are $\bU$ and $P$. The stress and heat flux of the wave are functionals of $\bU$, whereas those of the particle are moments of $P$. The first three equations have the structure of the Navier--Stokes equations, although no constitutive relation has been assumed. For compatible initial and boundary data the system is equivalent to the kinetic equation (1) for every horizon, in the sense that a solution $(\bU,P)$ yields the solution $f=W+P$ of (1) and that $\bU-\avg{\bpsi(W+P)}$ remains constant in time.

\subsection{Continuous spectrum}\label{sec:spectrum}
Near equilibrium, what the wave and the particle carry can be stated explicitly. For a locally frozen relaxation time, the weight $e^{-\nu(s)}/\tau_s=e^{-s/\tau}/\tau$ has on $[0,\ell]$ the zeroth and first moments
\begin{equation}\label{eq:weights}
 \begin{aligned}
 &\int_0^\ell\frac{e^{-s/\tau}}{\tau}\,\dd s=1-e^{-\eta}=A_0(\eta),\\
 &\int_0^\ell s\,\frac{e^{-s/\tau}}{\tau}\,\dd s=\tau\big[1-(1+\eta)e^{-\eta}\big]=\tau A_1(\eta),
 \end{aligned}
\end{equation}
and the expansion $M_s=M-s\,\cD M+\mathcal O(s^2)$ turns (9) into
\begin{equation}\label{eq:wave_expansion}
 W=A_0M-A_1\,\tau\big(\partial_tM+\bxi\cdot\nabla_{\bx}M\big)+\mathcal O(\tau^2),
\end{equation}
with a remainder bounded uniformly in $\eta\ge0$. To compare the wave with the kinetic solution, let $\tau=\varepsilon\bar\tau$ and $\cD^{(0)}=\partial_t^{(0)}+\bxi\cdot\nabla_{\bx}$, where $\partial_t^{(0)}$ is the time derivative given by the Euler equations. Subtracting (23) from the Chapman--Enskog expansion of $f$ and using $1-A_0=e^{-\eta}$ and $1-A_1=(1+\eta)e^{-\eta}$, one obtains through first order
\begin{equation}\label{eq:wp_first_order}
 \begin{aligned}
 &W=A_0M-\varepsilon A_1\bar\tau\,\cD^{(0)}M,\\
 &P=e^{-\eta}M+\varepsilon\big[r^{(1)}-(1+\eta)e^{-\eta}\,\bar\tau\,\cD^{(0)}M\big],
 \end{aligned}
\end{equation}
where $r^{(1)}=f^{(1)}+\bar\tau\,\cD^{(0)}M$ is the first-order response to the remainder. Indeed, at order $\varepsilon^0$, (4) reads $\cD^{(0)}M=-f^{(1)}/\bar\tau+R^{(0)}$, whence $r^{(1)}=\bar\tau R^{(0)}$ with the leading-order remainder $R^{(0)}$.

The moments of (24) are most clearly expressed through three fluxes. Since $\avg{\bpsi\,\cD^{(0)}M}=\boldsymbol0$ by the Euler equations, the components share the conservative state as $\bU_W=A_0\bU$ and $\bU_P=e^{-\eta}\bU$, so that $\bj_W=\bj_P=\boldsymbol0$, $p_W=A_0p$ and $p_P=e^{-\eta}p$ at this order. The fluxes of $M$, of $-\varepsilon\bar\tau\cD^{(0)}M$ and of $\varepsilon r^{(1)}$ are, respectively,
\begin{equation}\label{eq:ns_split}
 \begin{aligned}
 &\bF_{\rm E}=\big(\rho\bu,\ \rho\bu\bu^T+p\bI,\ (\rho E+p)\bu\big) && \text{(Euler flux)},\\
 &\bF_{\rm R}=\big(\boldsymbol0,\ \stau,\ \stau\bu+\qtau\big) && \text{(relaxation flux)},\\
 &\bF_{\rm S}=\big(\boldsymbol0,\ \sNS-\stau,\ (\sNS-\stau)\bu+\qNS-\qtau\big) && \text{(correction flux)},
 \end{aligned}
\end{equation}
where
\begin{equation}
 \sNS=-\mu\big[\nabla_{\bx}\bu+(\nabla_{\bx}\bu)^T-\tfrac23(\nabla_{\bx}\cdot\bu)\bI\big],\qquad \qNS=-\kappa\nabla_{\bx}T
\end{equation}
are the Navier--Stokes stress and heat flux of the model, while $\stau$ and $\qtau$ denote the same laws formed with the relaxation viscosity $\tau p$ and the relaxation conductivity $c_p\tau p$, whose Prandtl number is one. Consequently, for every locally frozen $\eta$,
\begin{equation}\label{eq:spectrum}
 \bF_W=A_0\bF_{\rm E}+A_1\bF_{\rm R},\qquad
 \bF_P=e^{-\eta}\bF_{\rm E}+(1+\eta)e^{-\eta}\bF_{\rm R}+\bF_{\rm S},
\end{equation}
and the sum $\bF_W+\bF_P=\bF_{\rm E}+\bF_{\rm R}+\bF_{\rm S}$ is the Navier--Stokes flux of the model, whatever the value of $\eta$. When $\tau=\mu/p$, the relaxation term reproduces the full viscous stress, $\stau=\sNS$, and the fraction $\Pr$ of the heat flux, $\qtau=-c_p\mu\nabla_{\bx}T=\Pr\,\qNS$, so that
\begin{equation}\label{eq:ns_split_mu}
 \bF_{\rm R}=\big(\boldsymbol0,\ \sNS,\ \sNS\bu+\Pr\,\qNS\big),\qquad
 \bF_{\rm S}=\big(\boldsymbol0,\ \boldsymbol0,\ (1-\Pr)\,\qNS\big),
\end{equation}
and the stress and heat flux in (20) become
\begin{equation}\label{eq:spectrum_sigma}
 \begin{aligned}
 &\boldsymbol\sigma_W=A_1\,\sNS, && \bq_W=A_1\Pr\,\qNS,\\
 &\boldsymbol\sigma_P=(1+\eta)e^{-\eta}\,\sNS, && \bq_P=(1+\eta)e^{-\eta}\Pr\,\qNS+(1-\Pr)\,\qNS .
 \end{aligned}
\end{equation}
The weights are listed in Table~1 and plotted in Fig.~2.

The two ends of this family are the kinetic equation and the Navier--Stokes equations. As $\eta\to0$, the wave vanishes, $\cS_\ell\to M/\tau$ and the particle carries all three fluxes, so that (20) reduces to its fourth equation, which is (1) for $f=P$. As $\eta\to\infty$, the particle reduces to $P=\varepsilon r^{(1)}+\mathcal O(\varepsilon^2)$. The wave then carries $\bF_{\rm E}+\bF_{\rm R}$ and the particle only $\bF_{\rm S}$, and the first three equations of (20) become the Navier--Stokes equations of the model,
\begin{equation}
 \partial_t\bU+\nabla_{\bx}\cdot\big(\bF_{\rm E}+\bF_{\rm R}+\bF_{\rm S}\big)=\boldsymbol0,
\end{equation}
in which the last flux is supplied by the particle. Between these limits the fluxes are divided according to (27). On a mesh the ratio $\eta$ takes a different value in every cell, so that the passage from one description to the other needs neither a domain decomposition nor a switching criterion. The family of equivalent systems (20), parametrized by $\eta$, is what we call the continuous spectrum of wave--particle equations.
\begin{figure}[tbp]
\centering
\includegraphics[width=0.6\textwidth]{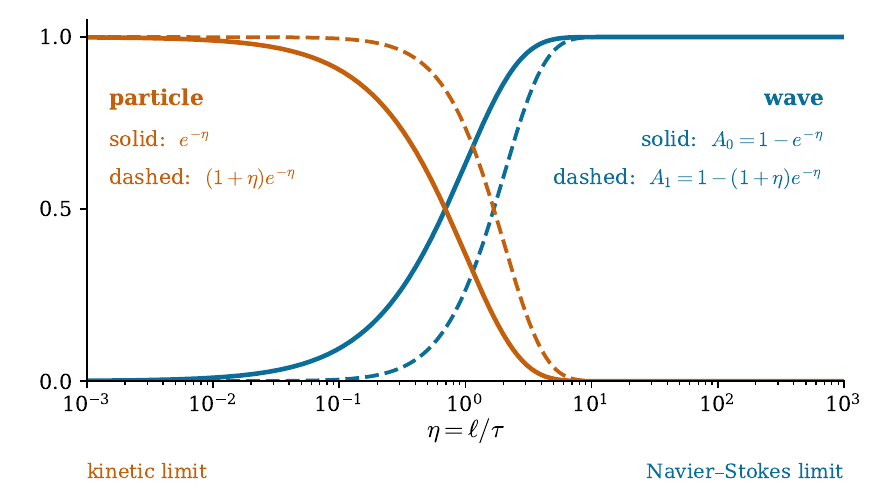}
\caption{Horizon weights against the horizon-to-relaxation ratio $\eta=\ell/\tau$. The shares of the Maxwellian and of the Euler flux carried by the wave and by the particle, $A_0$ and $e^{-\eta}$, are drawn solid, and the shares of the relaxation flux $\bF_{\rm R}$, $A_1$ and $(1+\eta)e^{-\eta}$, dashed.}\label{fig:weights}
\end{figure}
\begin{table}[tbp]
\centering\small
\caption{Horizon weights}\label{tab:weights}
\begin{tabular}{@{}lcccc@{}}
\toprule
$\eta$ & $A_0=1-e^{-\eta}$ & $A_1=1-(1+\eta)e^{-\eta}$ & $e^{-\eta}$ & $(1+\eta)e^{-\eta}$\\
\midrule
$0.1$ & $0.0952$ & $0.0047$ & $0.9048$ & $0.9953$\\
$1$ & $0.6321$ & $0.2642$ & $0.3679$ & $0.7358$\\
$4$ & $0.9817$ & $0.9084$ & $0.0183$ & $0.0916$\\
$10$ & $1-4.5\times10^{-5}$ & $0.9995$ & $4.5\times10^{-5}$ & $5.0\times10^{-4}$\\
\bottomrule
\end{tabular}
\tablelegend{$A_0$ and $e^{-\eta}$ are the shares of $M$ and $\bF_{\rm E}$ carried by the wave and the particle, and $A_1$ and $(1+\eta)e^{-\eta}$ are the shares of $\bF_{\rm R}$.}
\end{table}

\subsection{Specific models}\label{sec:models_theory}
The wave operator (9), the wave equation (13) and the macroscopic equations (17) and (19) are common to all collision models of a gas. The models differ in the remainder and, through it, in the particle, in the particle equation, in the response $r^{(1)}$ and in the correction flux $\bF_{\rm S}$. Five gas models are considered first, followed by two transport systems of a different kind with the same structure, neutron transport and radiative transfer.

\subsubsection{BGK model}\label{sec:bgk}
The BGK model~[38],
\begin{equation}
 \partial_tf+\bxi\cdot\nabla_{\bx}f=\frac{M-f}{\tau},
\end{equation}
with $\tau=\mu/p$, $\kappa=c_p\mu$ and $\Pr=1$, is the degenerate case of the decomposition. When the relaxation time of the model is used in (4), the remainder vanishes, and the integral solution (8) consists of the wave and the particle
\begin{equation}
 P=e^{-\nu(\ell)}f_\ell\ \ge0,
\end{equation}
the fraction of the upstream distribution that arrives from beyond the horizon without a collision. This is the original decomposition~[35], and with a horizon equal to the time step it is the wave--particle split of the unified gas-kinetic wave--particle method~[24] (Section~4.3). The particle equation,
\begin{equation}
 \partial_tP+\bxi\cdot\nabla_{\bx}P=-\frac{P}{\tau}+\cS_\ell=\frac{P_e-P}{\tau},
\end{equation}
contains no collision term. The particle is attenuated at the rate $1/\tau$ and fed by the horizon exchange. The wave--particle multiscale equations are
\begin{equation}
 \left\{
 \begin{aligned}
 &\partial_t\rho+\nabla_{\bx}\cdot(\rho\bu)=0,\\
 &\partial_t(\rho\bu)+\nabla_{\bx}\cdot\big(\rho\bu\bu^T+p\bI+\boldsymbol\sigma_W+\boldsymbol\sigma_P\big)=\boldsymbol0,\\
 &\partial_t(\rho E)+\nabla_{\bx}\cdot\big[(\rho E+p)\bu+(\boldsymbol\sigma_W+\boldsymbol\sigma_P)\bu+\bq_W+\bq_P\big]=0,\\
 &\partial_tP+\bxi\cdot\nabla_{\bx}P=\frac{P_e-P}{\tau}.
 \end{aligned}
 \right.
\end{equation}
In the continuous spectrum one has $f^{(1)}=-\bar\tau\cD^{(0)}M$ and $r^{(1)}=0$, so that
\begin{equation}
 \begin{aligned}
 &W=A_0M-\varepsilon A_1\bar\tau\,\cD^{(0)}M,\\
 &P=e^{-\eta}M-\varepsilon(1+\eta)e^{-\eta}\,\bar\tau\,\cD^{(0)}M,
 \end{aligned}
\end{equation}
\begin{equation}
 \bF_W=A_0\bF_{\rm E}+A_1\bF_{\rm R},\qquad \bF_P=e^{-\eta}\bF_{\rm E}+(1+\eta)e^{-\eta}\bF_{\rm R},
\end{equation}
with $\bF_{\rm R}=(\boldsymbol0,\ \sNS,\ \sNS\bu+\qNS)$ and $\bF_{\rm S}=\boldsymbol0$, that is,
\begin{equation}
 \begin{aligned}
 &\boldsymbol\sigma_W=A_1\,\sNS, && \bq_W=A_1\,\qNS,\\
 &\boldsymbol\sigma_P=(1+\eta)e^{-\eta}\,\sNS, && \bq_P=(1+\eta)e^{-\eta}\,\qNS .
 \end{aligned}
\end{equation}
All moments of the particle decay exponentially with $\eta$. Among the gas models considered here, the BGK model is the only one in which the particle is exponentially small in the continuum regime. In the others it remains of first order in $\varepsilon$, because it has to supply the correction flux.

\subsubsection{Shakhov model}\label{sec:shakhov}
The Shakhov model~[39] corrects the Prandtl number of the BGK model through the heat flux of the relaxation target,
\begin{equation}
 \begin{aligned}
 &\partial_tf+\bxi\cdot\nabla_{\bx}f=\frac{S-f}{\tau},\\
 &S=M\left[1+(1-\Pr)\frac{\bc\cdot\bq}{5pT}\left(\frac{|\bc|^2}{T}-5\right)\right],
 \end{aligned}
\end{equation}
where $\tau=\mu/p$, $\bq=\frac12\avg{|\bc|^2\bc f}$ is the heat flux and $\kappa=c_p\mu/\Pr$. The target has the conservative moments of $f$, no stress and the heat flux $(1-\Pr)\bq$. Splitting it as $S=M+(S-M)$ gives
\begin{equation}
 \begin{aligned}
 &R=\frac{S-M}{\tau},\\
 &P=e^{-\nu(\ell)}f_\ell+\int_0^\ell\frac{(S-M)_s}{\tau_s}e^{-\nu(s)}\,\dd s .
 \end{aligned}
\end{equation}
The remainder is explicit, since it depends on $f$ only through $\bU$ and $\bq$, and the Prandtl-number correction of the model is carried entirely by the particle. One could instead let the wave integrate the full target and so make the remainder vanish. The target $S$, however, contains the heat flux, a moment of $f=W+P$ that obeys no conservation law. The wave would then cease to be a functional of $\bU$, and the conservation law in (20) would no longer close. The particle equation is
\begin{equation}
 \partial_tP+\bxi\cdot\nabla_{\bx}P=\frac{S-M}{\tau}-\frac{P}{\tau}+\cS_\ell,
\end{equation}
and the wave--particle multiscale equations are
\begin{equation}
 \left\{
 \begin{aligned}
 &\partial_t\rho+\nabla_{\bx}\cdot(\rho\bu)=0,\\
 &\partial_t(\rho\bu)+\nabla_{\bx}\cdot\big(\rho\bu\bu^T+p\bI+\boldsymbol\sigma_W+\boldsymbol\sigma_P\big)=\boldsymbol0,\\
 &\partial_t(\rho E)+\nabla_{\bx}\cdot\big[(\rho E+p)\bu+(\boldsymbol\sigma_W+\boldsymbol\sigma_P)\bu+\bq_W+\bq_P\big]=0,\\
 &\partial_tP+\bxi\cdot\nabla_{\bx}P=\frac{S-M}{\tau}+\frac{P_e-P}{\tau}.
 \end{aligned}
 \right.
\end{equation}
In the continuous spectrum, $f^{(1)}=S^{(1)}-\bar\tau\cD^{(0)}M$ and $r^{(1)}=S^{(1)}$, where $\varepsilon S^{(1)}$ is $S-M$ evaluated with the first-order heat flux,
\begin{equation}
 \varepsilon S^{(1)}=M\,(1-\Pr)\frac{\bc\cdot\qNS}{5pT}\left(\frac{|\bc|^2}{T}-5\right).
\end{equation}
Hence
\begin{equation}
 \begin{aligned}
 &W=A_0M-\varepsilon A_1\bar\tau\,\cD^{(0)}M,\\
 &P=e^{-\eta}M+\varepsilon\big[S^{(1)}-(1+\eta)e^{-\eta}\,\bar\tau\,\cD^{(0)}M\big],
 \end{aligned}
\end{equation}
\begin{equation}
 \bF_W=A_0\bF_{\rm E}+A_1\bF_{\rm R},\qquad \bF_P=e^{-\eta}\bF_{\rm E}+(1+\eta)e^{-\eta}\bF_{\rm R}+\bF_{\rm S},
\end{equation}
with $\bF_{\rm R}$ and $\bF_{\rm S}$ given by (28), the latter being the flux of $\varepsilon S^{(1)}$, and
\begin{equation}
 \begin{aligned}
 &\boldsymbol\sigma_W=A_1\,\sNS, && \bq_W=A_1\Pr\,\qNS,\\
 &\boldsymbol\sigma_P=(1+\eta)e^{-\eta}\,\sNS, && \bq_P=(1+\eta)e^{-\eta}\Pr\,\qNS+(1-\Pr)\,\qNS .
 \end{aligned}
\end{equation}
For $\Pr=2/3$, the continuum limit assigns the whole stress and two thirds of the heat flux to the wave and the remaining third of the heat flux to the particle, which tends to $\varepsilon S^{(1)}$.

\subsubsection{ES-BGK model}\label{sec:es}
The ellipsoidal statistical model~[40] corrects the Prandtl number through the stress of the target instead. The distribution relaxes towards an anisotropic Gaussian,
\begin{equation}
 \begin{aligned}
 &\partial_tf+\bxi\cdot\nabla_{\bx}f=\frac{G-f}{\tau},\\
 &G=\frac{\rho}{\sqrt{\det(2\pi\cT)}}\exp\Big(-\frac12\bc^T\cT^{-1}\bc\Big),\\
 &\cT=\frac{1}{\Pr}\,T\bI+\Big(1-\frac{1}{\Pr}\Big)\boldsymbol\Theta,
 \end{aligned}
\end{equation}
where $\tau=\mu/(\Pr\,p)$, $\rho\boldsymbol\Theta=\avg{\bc\bc^Tf}$ is the pressure tensor and $\kappa=c_p\mu/\Pr$. The Gaussian has the conservative moments of $f$ and no heat flux, and its stress is $(1-1/\Pr)$ times that of $f$. The split $G=M+(G-M)$ gives
\begin{equation}
 \begin{aligned}
 &R=\frac{G-M}{\tau},\\
 &P=e^{-\nu(\ell)}f_\ell+\int_0^\ell\frac{(G-M)_s}{\tau_s}e^{-\nu(s)}\,\dd s .
 \end{aligned}
\end{equation}
As for the Shakhov model, the remainder is explicit, since it depends on $f$ only through $\bU$ and the pressure tensor, and the wave cannot integrate $G$, because $\boldsymbol\Theta$ is not determined by $\bU$. The particle equation is
\begin{equation}
 \partial_tP+\bxi\cdot\nabla_{\bx}P=\frac{G-M}{\tau}-\frac{P}{\tau}+\cS_\ell,
\end{equation}
and the wave--particle multiscale equations are
\begin{equation}
 \left\{
 \begin{aligned}
 &\partial_t\rho+\nabla_{\bx}\cdot(\rho\bu)=0,\\
 &\partial_t(\rho\bu)+\nabla_{\bx}\cdot\big(\rho\bu\bu^T+p\bI+\boldsymbol\sigma_W+\boldsymbol\sigma_P\big)=\boldsymbol0,\\
 &\partial_t(\rho E)+\nabla_{\bx}\cdot\big[(\rho E+p)\bu+(\boldsymbol\sigma_W+\boldsymbol\sigma_P)\bu+\bq_W+\bq_P\big]=0,\\
 &\partial_tP+\bxi\cdot\nabla_{\bx}P=\frac{G-M}{\tau}+\frac{P_e-P}{\tau}.
 \end{aligned}
 \right.
\end{equation}
In the continuous spectrum, $f^{(1)}=G^{(1)}-\bar\tau\cD^{(0)}M$ and $r^{(1)}=G^{(1)}$, where $\varepsilon G^{(1)}$ is $G-M$ linearized about the Navier--Stokes stress,
\begin{equation}
 \varepsilon G^{(1)}=M\,\Big(1-\frac1\Pr\Big)\frac{\bc^T\sNS\,\bc}{2pT}.
\end{equation}
The relaxation time of this model is $\mu/(\Pr\,p)$ rather than $\mu/p$. The relaxation viscosity is therefore $\tau p=\mu/\Pr$ and the relaxation conductivity $c_p\tau p=\kappa$, so that $\stau=\sNS/\Pr$ and $\qtau=\qNS$, and (25) gives
\begin{equation}
 \begin{aligned}
 &\bF_{\rm R}=\Big(\boldsymbol0,\ \frac{1}{\Pr}\sNS,\ \frac{1}{\Pr}\sNS\bu+\qNS\Big),\\
 &\bF_{\rm S}=\Big(\boldsymbol0,\ \Big(1-\frac1\Pr\Big)\sNS,\ \Big(1-\frac1\Pr\Big)\sNS\bu\Big).
 \end{aligned}
\end{equation}
The first-order forms (24) and the fluxes (27) hold with these expressions, and
\begin{equation}
 \begin{aligned}
 &\boldsymbol\sigma_W=\frac{A_1}{\Pr}\,\sNS, && \bq_W=A_1\,\qNS,\\
 &\boldsymbol\sigma_P=\frac{(1+\eta)e^{-\eta}}{\Pr}\,\sNS+\Big(1-\frac1\Pr\Big)\sNS, && \bq_P=(1+\eta)e^{-\eta}\,\qNS .
 \end{aligned}
\end{equation}
The Shakhov and ES-BGK models thus correct the Prandtl number in complementary ways. In the Shakhov model the wave carries the full stress and the fraction $\Pr$ of the heat flux, and the particle supplies the missing heat flux. In the ES-BGK model the wave carries the full heat flux and the stress $\sNS/\Pr$, and the particle supplies the stress $(1-1/\Pr)\sNS$, which for $\Pr<1$ is opposite in sign to $\sNS$.

\subsubsection{Boltzmann equation}\label{sec:boltzmann}
The Boltzmann equation reads
\begin{equation}
 \begin{aligned}
 &\partial_tf+\bxi\cdot\nabla_{\bx}f=\frac{1}{\varepsilon}Q(f,f),\\
 &Q(f,f)=\int_{\mathbb R^3}\int_{S^2}B(|\bg|,\theta)\big(f'f'_*-ff_*\big)\,\dd\bn\,\dd\bxi_* ,
 \end{aligned}
\end{equation}
with $\bg=\bxi-\bxi_*$, $\cos\theta=\bg\cdot\bn/|\bg|$, $\bxi'=(\bxi+\bxi_*)/2+|\bg|\bn/2$, $\bxi'_*=\bxi+\bxi_*-\bxi'$ and a collision kernel $B$ independent of $\varepsilon$. Here the relaxation time is not part of the model. The polarization $Q(g,h)=\frac12[Q(g+h,g+h)-Q(g,g)-Q(h,h)]$ defines the linearization $L_M\delta f=2Q(M,\delta f)$, and the first-order Chapman--Enskog correction, the solution of $L_Mf^{(1)}=\cD^{(0)}M$ with $\avg{\bpsi f^{(1)}}=\boldsymbol0$, determines $\mu$ and $\kappa$. With $\tau=\mu/p$,
\begin{equation}
 \begin{aligned}
 &R=\frac{1}{\varepsilon}Q(f,f)-\frac{M-f}{\tau},\\
 &P=e^{-\nu(\ell)}f_\ell+\int_0^\ell R_se^{-\nu(s)}\,\dd s .
 \end{aligned}
\end{equation}
The relaxation time parametrizes the decomposition but enters neither the Boltzmann equation nor its solutions. The particle equation is
\begin{equation}
 \partial_tP+\bxi\cdot\nabla_{\bx}P=\frac{1}{\varepsilon}Q(W+P,W+P)-\frac{M-W}{\tau}+\cS_\ell,
\end{equation}
and the wave--particle multiscale equations are
\begin{equation}
 \left\{
 \begin{aligned}
 &\partial_t\rho+\nabla_{\bx}\cdot(\rho\bu)=0,\\
 &\partial_t(\rho\bu)+\nabla_{\bx}\cdot\big(\rho\bu\bu^T+p\bI+\boldsymbol\sigma_W+\boldsymbol\sigma_P\big)=\boldsymbol0,\\
 &\partial_t(\rho E)+\nabla_{\bx}\cdot\big[(\rho E+p)\bu+(\boldsymbol\sigma_W+\boldsymbol\sigma_P)\bu+\bq_W+\bq_P\big]=0,\\
 &\partial_tP+\bxi\cdot\nabla_{\bx}P=\frac{1}{\varepsilon}Q(W+P,W+P)-\frac{M-W-P}{\tau}+\frac{P_e-P}{\tau}.
 \end{aligned}
 \right.
\end{equation}
Since the exchange terms cancel when the wave and particle equations are added, the full collision operator is retained for every horizon, including the wave--wave, cross and particle--particle collisions in $Q(W+P,W+P)=Q(W,W)+2Q(W,P)+Q(P,P)$. In the continuous spectrum, $r^{(1)}=f^{(1)}+\bar\tau\cD^{(0)}M$ and
\begin{equation}
 \begin{aligned}
 &W=A_0M-\varepsilon A_1\bar\tau\,\cD^{(0)}M,\\
 &P=e^{-\eta}M+\varepsilon\big[f^{(1)}+A_1\bar\tau\,\cD^{(0)}M\big],
 \end{aligned}
\end{equation}
\begin{equation}
 \bF_W=A_0\bF_{\rm E}+A_1\bF_{\rm R},\qquad \bF_P=e^{-\eta}\bF_{\rm E}+(1+\eta)e^{-\eta}\bF_{\rm R}+\bF_{\rm S}.
\end{equation}
For a general $\tau$ the fluxes are those of (25), and the stress and heat flux of the particle tend to $\sNS-\stau$ and $\qNS-\qtau$ as $\eta\to\infty$. With $\tau=\mu/p$ the relaxation term reproduces the whole Navier--Stokes stress, and
\begin{equation}
 \begin{aligned}
 &\boldsymbol\sigma_W=A_1\,\sNS, && \bq_W=-A_1\,c_p\mu\,\nabla_{\bx}T=A_1\Pr\,\qNS,\\
 &\boldsymbol\sigma_P=(1+\eta)e^{-\eta}\,\sNS, && \bq_P=(1+\eta)e^{-\eta}\Pr\,\qNS-(\kappa-c_p\mu)\nabla_{\bx}T,
 \end{aligned}
\end{equation}
where $-(\kappa-c_p\mu)\nabla_{\bx}T=(1-\Pr)\qNS$ and $\Pr$ is the Prandtl number of the kernel, close to $2/3$. In the continuum limit the wave thus carries the Euler flux together with the relaxation part of the viscous flux, whose Prandtl number is one, and the particle corrects the heat conductivity to that of the Boltzmann operator.

\subsubsection{Landau equation}\label{sec:landau}
The Landau equation for Coulomb collisions, written here without a force field,
\begin{equation}
 \begin{aligned}
 &\partial_tf+\bxi\cdot\nabla_{\bx}f=\frac{1}{\varepsilon}Q_L(f,f),\\
 &Q_L(f,f)=\nabla_{\bxi}\cdot\int_{\mathbb R^3}\Phi(\bxi-\bxi_*)\big[f_*\nabla_{\bxi}f-f\,\nabla_{\bxi_*}f_*\big]\dd\bxi_*,
 \end{aligned}
\end{equation}
with $\Phi(\boldsymbol z)=|\boldsymbol z|^{-1}(\bI-\boldsymbol z\boldsymbol z^T/|\boldsymbol z|^2)$, shows that the decomposition does not rely on the collision term being an integral operator. The operator $Q_L$ conserves mass, momentum and energy, vanishes exactly on Maxwellians and has a Navier--Stokes limit, with a viscosity $\mu_L$ and a conductivity $\kappa_L$ determined by its linearization. The three properties required in Section~2 therefore hold, and the equations above apply without change. With $\tau=\mu_L/p$,
\begin{equation}
 \begin{aligned}
 &R=\frac{1}{\varepsilon}Q_L(f,f)-\frac{M-f}{\tau},\\
 &P=e^{-\nu(\ell)}f_\ell+\int_0^\ell R_se^{-\nu(s)}\,\dd s,
 \end{aligned}
\end{equation}
the particle equation is
\begin{equation}
 \partial_tP+\bxi\cdot\nabla_{\bx}P=\frac{1}{\varepsilon}Q_L(W+P,W+P)-\frac{M-W}{\tau}+\cS_\ell,
\end{equation}
and the wave--particle multiscale equations are
\begin{equation}
 \left\{
 \begin{aligned}
 &\partial_t\rho+\nabla_{\bx}\cdot(\rho\bu)=0,\\
 &\partial_t(\rho\bu)+\nabla_{\bx}\cdot\big(\rho\bu\bu^T+p\bI+\boldsymbol\sigma_W+\boldsymbol\sigma_P\big)=\boldsymbol0,\\
 &\partial_t(\rho E)+\nabla_{\bx}\cdot\big[(\rho E+p)\bu+(\boldsymbol\sigma_W+\boldsymbol\sigma_P)\bu+\bq_W+\bq_P\big]=0,\\
 &\partial_tP+\bxi\cdot\nabla_{\bx}P=\frac{1}{\varepsilon}Q_L(W+P,W+P)-\frac{M-W-P}{\tau}+\frac{P_e-P}{\tau}.
 \end{aligned}
 \right.
\end{equation}
In the continuous spectrum, with $f^{(1)}$ the first-order correction of $Q_L$, with $\sNS$ and $\qNS$ formed with $\mu_L$ and $\kappa_L$, and with $\Pr=c_p\mu_L/\kappa_L$,
\begin{equation}
 \begin{aligned}
 &W=A_0M-\varepsilon A_1\bar\tau\,\cD^{(0)}M,\\
 &P=e^{-\eta}M+\varepsilon\big[f^{(1)}+A_1\bar\tau\,\cD^{(0)}M\big],
 \end{aligned}
\end{equation}
\begin{equation}
 \begin{aligned}
 &\boldsymbol\sigma_W=A_1\,\sNS, && \bq_W=A_1\Pr\,\qNS,\\
 &\boldsymbol\sigma_P=(1+\eta)e^{-\eta}\,\sNS, && \bq_P=(1+\eta)e^{-\eta}\Pr\,\qNS+(1-\Pr)\,\qNS .
 \end{aligned}
\end{equation}

\subsubsection{Neutron transport}\label{sec:neutron}
The construction is not restricted to gases. The one-speed transport equation with isotropic scattering, the basic model of neutron transport, reads
\begin{equation}\label{eq:linear_transport}
 \frac1c\,\partial_t\psi+\bOm\cdot\nabla_{\bx}\psi+\sigma_t\psi=\frac{\sigma_s\phi+q}{4\pi},\qquad \phi=\int_{S^2}\psi\,\dd\bOm,
\end{equation}
with the angular flux $\psi(\bx,\bOm,t)$, the speed $c$, the total cross section $\sigma_t=\sigma_s+\sigma_a$ of scattering and absorption, and an isotropic source $q$. With $\bxi=c\bOm$, the relaxation time $\tau=1/(c\sigma_t)$, which is the mean free time, and the source function
\begin{equation}\label{eq:source_function}
 S=\frac{\sigma_s\phi+q}{4\pi\sigma_t},
\end{equation}
it takes the form (4), with $S$ in place of the Maxwellian and a vanishing remainder,
\begin{equation}
 \partial_t\psi+\bxi\cdot\nabla_{\bx}\psi=\frac{S-\psi}{\tau}.
\end{equation}
The scalar flux $\phi$ determines the source function and therefore takes the place of the conservative variables. The only collision invariant is the constant, which the relaxation conserves only in the absence of absorption and sources, and the conservation law is accordingly replaced by the balance law
\begin{equation}
 \frac1c\,\partial_t\phi+\nabla_{\bx}\cdot\bJ+\sigma_a\phi=q,\qquad \bJ=\int_{S^2}\bOm\,\psi\,\dd\bOm .
\end{equation}
The wave and the particle are
\begin{equation}
 W=\int_0^{\ell}\frac{S_s}{\tau_s}e^{-\nu(s)}\,\dd s,\qquad P=e^{-\nu(\ell)}\psi_\ell\ \ge0,
\end{equation}
the flux emitted by scattering and by the source within the horizon and the uncollided flux from beyond it. With the horizon length $L=c\ell$, the ratio $\eta=\ell/\tau=\sigma_tL$ is the optical thickness of the horizon. The wave and particle equations are (13) and (18) with $S$ in place of $M$ and $R=0$. In terms of the currents $\bJ_Z=\int\bOm\,Z\,\dd\bOm$ of the two components, the wave--particle multiscale equations for the unknowns $\phi$ and $P$ are
\begin{equation}
 \left\{
 \begin{aligned}
 &\frac1c\,\partial_t\phi+\nabla_{\bx}\cdot\big(\bJ_W+\bJ_P\big)+\sigma_a\phi=q,\\
 &\frac1c\,\partial_tP+\bOm\cdot\nabla_{\bx}P=\sigma_t\,(P_e-P).
 \end{aligned}
 \right.
\end{equation}
In the continuous spectrum the expansion (23) holds with $S$ in place of $M$. An isotropic distribution carries no current, so the counterpart of the Euler flux is absent, and
\begin{equation}\label{eq:currents}
 \begin{aligned}
 &\bJ_W=A_1\,\bJ_{\rm D},\qquad \bJ_P=(1+\eta)e^{-\eta}\,\bJ_{\rm D},\\
 &\bJ_{\rm D}=-\frac{1}{3\sigma_t}\nabla_{\bx}\big(4\pi S\big)=-\frac{1}{3\sigma_t}\nabla_{\bx}\frac{\sigma_s\phi+q}{\sigma_t},
 \end{aligned}
\end{equation}
where $\bJ_{\rm D}$ is Fick's law. In the diffusive scaling $4\pi S=\phi$ at leading order and $\bJ_{\rm D}=-\nabla_{\bx}\phi/(3\sigma_t)$. There is no correction flux. As $\eta\to0$ the system is the transport equation for $\psi=P$. As $\eta\to\infty$ the particle becomes exponentially small, as for the BGK model, and the balance law turns into the diffusion equation
\begin{equation}
 \frac1c\,\partial_t\phi-\nabla_{\bx}\cdot\Big(\frac{1}{3\sigma_t}\nabla_{\bx}\phi\Big)+\sigma_a\phi=q .
\end{equation}
With the horizon equal to the time step, the same integral solution underlies the wave--particle method for the neutron transport equation~[41].

\subsubsection{Radiative transfer}\label{sec:radiation}
Grey radiative transfer couples the radiation intensity $I(\bx,\bOm,t)$ to the material temperature $T$,
\begin{equation}
 \begin{aligned}
 &\frac1c\,\partial_tI+\bOm\cdot\nabla_{\bx}I+\sigma_tI=\sigma_aB(T)+\frac{\sigma_s\phi}{4\pi},\qquad \phi=\int_{S^2}I\,\dd\bOm,\\
 &C_v\,\partial_tT=\sigma_a\big(\phi-4\pi B(T)\big),
 \end{aligned}
\end{equation}
where $c$ is the speed of light, $B(T)=a_RcT^4/(4\pi)$ the Planck emission with the radiation constant $a_R$, $\sigma_a$ and $\sigma_s$ the absorption and scattering coefficients, $\sigma_t=\sigma_a+\sigma_s$, and $C_v$ the heat capacity of the material. With $\bxi=c\bOm$, $\tau=1/(c\sigma_t)$ and the source function
\begin{equation}
 S=\frac{1}{\sigma_t}\Big(\sigma_aB(T)+\frac{\sigma_s\phi}{4\pi}\Big),
\end{equation}
the transfer equation again takes the form $\partial_tI+\bxi\cdot\nabla_{\bx}I=(S-I)/\tau$ with a vanishing remainder. The macroscopic state is now the pair $(\phi,T)$, which determines the source function, and the conserved quantity is the total energy $\phi/c+C_vT$. The wave is the radiation emitted thermally or scattered within the horizon, and the particle consists of the photons that arrive from beyond the horizon without interacting,
\begin{equation}
 W=\int_0^{\ell}\frac{S_s}{\tau_s}e^{-\nu(s)}\,\dd s,\qquad P=e^{-\nu(\ell)}I_\ell\ \ge0 .
\end{equation}
The wave--particle multiscale equations for the unknowns $\phi$, $T$ and $P$ are
\begin{equation}
 \left\{
 \begin{aligned}
 &\frac1c\,\partial_t\phi+\nabla_{\bx}\cdot\big(\bJ_W+\bJ_P\big)=\sigma_a\big(4\pi B(T)-\phi\big),\\
 &C_v\,\partial_tT=\sigma_a\big(\phi-4\pi B(T)\big),\\
 &\frac1c\,\partial_tP+\bOm\cdot\nabla_{\bx}P=\sigma_t\,(P_e-P),
 \end{aligned}
 \right.
\end{equation}
with the currents given by (72) for the present source function. In the optically thick regime, $\phi=4\pi B(T)=a_RcT^4$ at leading order, the particle is exponentially small, and the sum of the first two equations becomes the equilibrium diffusion equation
\begin{equation}
 \partial_t\big(C_vT+a_RT^4\big)=\nabla_{\bx}\cdot\Big(\frac{c}{3\sigma_t}\nabla_{\bx}\big(a_RT^4\big)\Big).
\end{equation}
Although the emission is nonlinear in the temperature, the structure of the decomposition is unchanged. With the horizon equal to the time step, the same integral solution underlies the unified gas-kinetic scheme for radiative transfer~[21] and the wave--particle method for photon transport~[26].
\section{Coupled WP iteration}\label{sec:algorithm}

\subsection{Architecture}
For steady flows, the multiscale equations (20) reduce to a macroscopic and a kinetic equation,
\begin{equation}\label{eq:steady}
 \begin{aligned}
 &\nabla_{\bx}\cdot\big(\bF_W[\bU]+\bF_P[P]\big)=\boldsymbol0,\\
 &\bxi\cdot\nabla_{\bx}P=R+\frac{P_e[\bU]-P}{\tau},
 \end{aligned}
\end{equation}
which are coupled in both directions. The particle enters the macroscopic equation through its flux, that is, through $\boldsymbol\sigma_P$ and $\bq_P$, and the conservative variables enter the kinetic equation through $W$, $M$, $\tau$ and $P_e$. The coupled WP iteration solves them in turn. One outer iteration, which leads from the accepted state $(\bU^k,P^k)$ to the next, has the structure
\begin{equation}\label{eq:scheme}
 \begin{aligned}
 (\bU^k,P^k)\ &\xrightarrow{\ \text{W iteration, }P^k\text{ fixed}\ }\ \widetilde\bU^{k+1},\ \widetilde W^{k+1},\ \widetilde P_e^{h,k+1}\\
 &\xrightarrow{\ \text{traction }\Theta^k;\ \text{P iteration, }\widetilde W^{k+1}\text{ fixed}\ }\ P\ \xrightarrow{\ \text{acceptance}\ }\ (\bU^{k+1},P^{k+1}).
 \end{aligned}
\end{equation}
In the W iteration the macroscopic equation is solved for $\bU$ with the particle, and hence its flux, frozen. Only the wave and its flux follow $\bU$, no collision term is evaluated, and the cost does not depend on the velocity grid. The result is a predicted state $\widetilde\bU^{k+1}$ together with its wave and its endpoint. The change of the endpoint is passed to the kinetic equation as a source, the endpoint traction $\Theta^k$. The P iteration then solves the kinetic equation for $P$ with the wave fixed at the predicted wave, and this is the only place where the collision term is evaluated. Finally, the acceptance redefines $\bU$ as the moment of $f=\widetilde W^{k+1}+P$ and $P$ as $f-W(\bU)$.

Throughout, the unknowns are $\bU$ and $P$. The wave is always evaluated from $\bU$, and $f=W+P$. The residuals are evaluated on the accepted state. All quantities introduced for preconditioning, namely the pseudo-time step, the approximate Jacobians, the rate $1/\widehat\tau$ and the numbers of inner iterations, multiply increments only and therefore leave the converged solution unaffected. The division of work between the two iterations follows $\eta$. Where $\eta\gg1$, the wave carries $\bF_{\rm E}+\bF_{\rm R}$, the W iteration amounts to a Navier--Stokes solve and the P iteration contributes a small correction. Where $\eta\ll1$, the wave vanishes, the W iteration has no effect and the P iteration is the conventional discrete-ordinate iteration.

\subsection{Discrete system and residuals}
Physical space is divided into finite-volume cells $\Omega_i$ with faces $\Gamma\subset\partial\Omega_i$ of area $|\Gamma|$ and outward normal $\bn_\Gamma$, the neighbour across $\Gamma$ being $i_\Gamma$. Velocity space is represented by nodes $\bxi_j$ with quadrature weights $\varpi_j$, and $\bpsi_j=\bpsi(\bxi_j)$. The discrete conservative variables are $\bU_i=\sum_j\varpi_j\bpsi_jf_{i,j}$, and $M_i$ denotes the discrete Maxwellian whose quadrature moments equal $\bU_i$. The horizon is taken as a cell-crossing time,
\begin{equation}\label{eq:horizon}
 \ell_i=C_\ell\,\frac{\Delta_i}{|\bu_i|+a_i},\qquad \eta_i=\frac{\ell_i}{\tau_i},
\end{equation}
with the cell size $\Delta_i$, the sound speed $a_i$ and a horizon number $C_\ell$ of order one, so that $\eta_i$ compares the cell size with the local mean free path.

Three ingredients make up the discrete system, the wave in the cells, the fluxes of the two components through the faces, and a compatibility between the kinetic and the macroscopic fluxes. The cell wave is the first-order truncation of (23),
\begin{equation}\label{eq:discrete_wave}
 W_i=A_0(\eta_i)M_i-A_1(\eta_i)\,\tau_i\big(\partial_tM+\bxi\cdot\nabla_{\bx}M\big)^h_i,
\end{equation}
in which the derivative of the Maxwellian is formed from the cell gradient of $\bU$ and the Euler equations~[42]. It is determined by $\bU$ and satisfies $\sum_j\varpi_j\bpsi_jW_{i,j}=A_0(\eta_i)\bU_i$. The wave flux is the second-order gas-kinetic flux~[42] of the reconstructed face state $\bU_\Gamma$, with its Euler part weighted by $A_0$ and its Navier--Stokes part by $A_1$,
\begin{equation}\label{eq:wave_flux}
 \bF^h_{W,\Gamma}=A_{0,\Gamma}\Big[\bF_{\rm E}(\bU_\Gamma)+\frac{\Delta t}{2}\,\partial_t\bF_{\rm E}(\bU_\Gamma)\Big]\bn_\Gamma+A_{1,\Gamma}\,\bF_{\rm R}(\bU_\Gamma)\,\bn_\Gamma,
\end{equation}
where $\Delta t$ is the averaging interval of the flux and $\bF_{\rm R}$ is evaluated with the viscosity $\tau_\Gamma p_\Gamma$ and the conductivity $c_p\tau_\Gamma p_\Gamma$. This flux is the discrete counterpart of $\bF_W=A_0\bF_{\rm E}+A_1\bF_{\rm R}$, and for $A_{0,\Gamma}=A_{1,\Gamma}=1$ it is the gas-kinetic Navier--Stokes flux. The same face distribution, evaluated at the velocity nodes and corrected so that its quadrature moments equal (83), defines the velocity-resolved wave transport $D^h_W(W)_{i,j}$. The particle is transported node by node by a second-order upwind discrete-ordinate flux, whose cell divergence is $D^h_P(P)_{i,j}$ and whose quadrature moment is $\bF^h_{P,\Gamma}$. By construction, at interior and boundary faces alike,
\begin{equation}\label{eq:flux_compatibility}
 \sum_j\varpi_j\bpsi_j\big[D^h_W(W)+D^h_P(P)\big]_{i,j}=\nabla_h\cdot\big(\bF^h_W+\bF^h_P\big)\big|_i
 =\frac{1}{|\Omega_i|}\sum_{\Gamma\subset\partial\Omega_i}|\Gamma|\big(\bF^h_{W,\Gamma}+\bF^h_{P,\Gamma}\big).
\end{equation}
The reconstruction, the limiters, the time averaging of the fluxes and the wall treatment are standard components of finite-volume and discrete-ordinate methods and are not discussed here. The iteration relies only on (82)--(84).

Let $C^h(f)$ denote the discrete collision term of the model, including the factor $1/\varepsilon$, with $\sum_j\varpi_j\bpsi_jC^h_j=\boldsymbol0$. The discrete remainder and the discrete endpoint are
\begin{equation}\label{eq:discrete_endpoint}
 R^h=C^h(f)-\frac{M-f}{\tau},\qquad P_e^h=M-W-\tau D^h_W(W).
\end{equation}
The endpoint satisfies $(M-W-P_e^h)/\tau=D^h_W(W)$ identically, which is the steady discrete form of the wave equation (13) with $\cS_\ell=P_e/\tau$. On a smooth near-equilibrium state, $P_e^h=e^{-\eta}M+\mathcal O(\varepsilon)$. The residuals of an accepted state $(\bU,P)$ with $f=W(\bU)+P$ are
\begin{equation}\label{eq:residuals}
 \begin{aligned}
 &R_U=-\nabla_h\cdot\big(\bF^h_W+\bF^h_P\big),\\
 &R_P=R^h+\frac{P_e^h-P}{\tau}-D^h_P(P)=C^h(f)-D^h_W(W)-D^h_P(P).
 \end{aligned}
\end{equation}
The first form of $R_P$ is the steady discrete form of the particle equation (18). The second follows by substituting (85), and it is the form in which the residual is evaluated. By (84) and the conservation property of $C^h$, the quadrature moments of $R_P$ equal $R_U$. The macroscopic residual is thus the conservative part of the particle residual, and the discrete steady system $R_U=\boldsymbol0$, $R_P=0$ is the discrete kinetic equation
\begin{equation}\label{eq:discrete_kinetic}
 D^h_W(W)+D^h_P(P)=C^h(f)
\end{equation}
for $f=W(\bU)+P$. The residuals are measured in the volume-weighted norms
\begin{equation}
 \|R_U\|=\Big(\sum_i|\Omega_i|\,|R_{U,i}|^2\Big)^{1/2},\qquad
 \|R_P\|=\Big(\sum_i|\Omega_i|\sum_j\varpi_jR_{P,i,j}^2\Big)^{1/2}.
\end{equation}

\subsection{W iteration}\label{sec:W_update}
The W iteration performs $N_U$ updates of $\bU$ with the particle fixed at $P^k$, so that the particle flux keeps its accepted value $\bF^{h,k}_P$. The $n$th update starts from $\bU^{(n-1)}$, with $\bU^{(0)}=\bU^k$, and from the residual
\begin{equation}\label{eq:W_residual}
 R_U^{(n-1)}=-\nabla_h\cdot\big(\bF^h_W(\bU^{(n-1)})+\bF^{h,k}_P\big),
\end{equation}
in which only the wave flux is recomputed. The conservation law is advanced by one pseudo-time step in delta form~[30],
\begin{equation}\label{eq:W_update}
 d_i\,\Delta\bU_i+\sum_{\Gamma\subset\partial\Omega_i}G_{i,\Gamma}\big(\Delta\bU_{i_\Gamma}\big)=|\Omega_i|\,R^{(n-1)}_{U,i} .
\end{equation}
The left-hand side is an approximate Jacobian. Its diagonal $d_i$ collects the pseudo-time term $|\Omega_i|/\Delta t_{\rm num}$ and the face spectral radii, and $G_{i,\Gamma}$ is the Euler-flux increment of the neighbour with a Rusanov-type dissipation. The right-hand side is the physical residual, which contains the full wave flux (83) and the frozen particle flux, and it alone determines the converged state. The system is solved approximately by $m_U$ forward--backward point-relaxation (PR) sweep pairs, a single pair being a lower--upper symmetric Gauss--Seidel (LU--SGS) sweep pair~[43], and the state is updated as
\begin{equation}\label{eq:U_update}
 \bU^{(n)}=\bU^{(n-1)}+\theta_U\,\Delta\bU,
\end{equation}
with the largest $\theta_U\in\{1,\frac12,\frac14,\dots\}$ for which $\rho>0$, $T>0$ and $W(\bU^{(n)})+P^k\ge0$. The wave, its flux and (89) are then evaluated anew, with $P^k$ unchanged. After $N_U$ updates, $\widetilde\bU^{k+1}=\bU^{(N_U)}$ defines the predicted wave $\widetilde W^{k+1}=W(\widetilde\bU^{k+1})$ and, through (85), the predicted endpoint $\widetilde P_e^{h,k+1}$. In the absence of a particle and with $A_{0,\Gamma}=A_{1,\Gamma}=1$, the W iteration is a gas-kinetic Navier--Stokes solver on the same mesh.

\subsection{P iteration}\label{sec:P_update}
For finite $\eta$ the particle carries the share $e^{-\eta}M$ of the Maxwellian, $\bU_P=e^{-\eta}\bU$. The change of $\bU$ predicted by the W iteration must therefore reach the particle before the particle equation is solved, and this is the role of the endpoint traction
\begin{equation}\label{eq:traction}
 \Theta^k=\frac{\widetilde P_e^{h,k+1}-P_e^{h,k}}{\widehat\tau^k}.
\end{equation}
The traction is the change of the endpoint scaled by the rate $1/\widehat\tau$, which is either $1/\tau$ or a bound of the stiffest rate of the discrete collision term (Section~3.6). Because the same rate multiplies the increment of the particle in the update below, the particle follows the predicted change of its equilibrium share, $\Delta P=\Delta(e^{-\eta}M)+\mathcal O(\varepsilon)$ near equilibrium. The traction vanishes at convergence. It should not be confused with the relaxation $(P_e^h-P)/\tau$ contained in $R_P$, which belongs to the discrete equation and is balanced, not removed, in a steady state.

The P iteration performs $N_P$ updates of $P$ with the wave fixed at $\widetilde W^{k+1}$. The $n$th update starts from $P^{k,n-1}$, with $P^{k,0}=P^k$, and from $f^{k,n-1}=\widetilde W^{k+1}+P^{k,n-1}$, and it solves
\begin{equation}\label{eq:P_update}
 \begin{aligned}
 &\Big[\Big(\frac{1}{\Delta t_{\rm num}}+\frac{1}{\widehat\tau}\Big)\bI+D^h_{\rm up}\Big]\Delta P=b^{(n)},\\
 &b^{(1)}=R_P^k+\Theta^k,\qquad
 b^{(n)}=C^h\big(f^{k,n-1}\big)-D^h_W\big(\widetilde W^{k+1}\big)-D^h_P\big(P^{k,n-1}\big)\quad(n\ge2),
 \end{aligned}
\end{equation}
where $D^h_{\rm up}$ is the first-order upwind transport operator acting on increments. The later right-hand sides are the residual (86) with the wave fixed at $\widetilde W^{k+1}$. They need no endpoint, so that the traction enters the first update only. When $\widehat\tau=\tau$, the first right-hand side is
\begin{equation}\label{eq:rhs_meaning}
 b^{(1)}=R^h(f^k)+\frac{\widetilde P_e^{h,k+1}-P^k}{\tau}-D^h_P(P^k),
\end{equation}
which shows that the traction simply replaces the accepted endpoint in $R_P^k$ by the predicted one. System (93) is solved independently for every velocity node by $m_P$ LU--SGS sweep pairs over the cells. Since the right-hand side contains the second-order particle flux, the first-order operator on the left preconditions the increment without changing the converged solution. The particle is updated as $P^{k,n}=P^{k,n-1}+\theta\,\Delta P$ with the largest cell-wise factor $0<\theta\le1$ for which $f^{k,n}=\widetilde W^{k+1}+P^{k,n}$ remains non-negative, so that positivity is imposed on $f$ and not on the signed particle. The collision term, together with $\widehat\tau$ if it depends on $f$, is then evaluated on $f^{k,n}$.

After the $N_P$ updates, the distribution $f^{k+1}=\widetilde W^{k+1}+P^{k,N_P}$ defines the accepted state,
\begin{equation}\label{eq:acceptance}
 \bU^{k+1}=\sum_j\varpi_j\bpsi_jf^{k+1}_j,\qquad P^{k+1}=f^{k+1}-W(\bU^{k+1}).
\end{equation}
The acceptance leaves $f$ unchanged and restores $\bU=\sum_j\varpi_j\bpsi_j(W+P)_j$. Because the accepted conservative variables are the moments of the corrected distribution rather than the prediction, the P iteration feeds back on the macroscopic state, and the two-way coupling is closed.

\subsection{Complete iteration}
The complete iteration starts from an initial field $\bU^0$, with the particle $P^0=M^0-W(\bU^0)$ and with $C^h$ evaluated on $f^0=M^0$, and repeats the following steps for $k=0,1,2,\dots$ until both residuals are below the tolerance.
\begin{enumerate}
\item On the accepted state $(\bU^k,P^k)$, with $f^k=W(\bU^k)+P^k$, the residuals $R_U^k$ and $R_P^k$, the endpoint $P_e^{h,k}$ and the rate $\widehat\tau^k$ are formed, and the iteration stops if $\|R_U\|$ and $\|R_P\|$ are below the tolerance.
\item The W iteration performs $N_U$ updates (90)--(91) with $P=P^k$ fixed, each with $m_U$ PR pairs, and yields $\widetilde\bU^{k+1}$, $\widetilde W^{k+1}$ and $\widetilde P_e^{h,k+1}$.
\item The endpoint traction $\Theta^k$ is formed by (92).
\item The P iteration performs $N_P$ updates (93) with $W=\widetilde W^{k+1}$ fixed, each with $m_P$ LU--SGS pairs, the positivity-limited update and one evaluation of $C^h$.
\item The acceptance (95) defines $\bU^{k+1}$ and $P^{k+1}$.
\end{enumerate}
The iteration is controlled by the numbers $N_U$ and $N_P$ of updates and the numbers $m_U$ and $m_P$ of sweep pairs, and a configuration is written as W$N_U$/PR$m_U$/P$N_P$. In rarefied flows a single macroscopic update, $N_U=1$, suffices. In the continuum regime $N_U$ ranges from 24 to 256 with $m_U$ between 2 and 6, while $N_P=1$ and $m_P=1$ throughout. One outer iteration thus costs $N_U$ macroscopic updates, which are independent of the velocity grid, and $N_P$ kinetic updates with one evaluation of the collision term each, so that additional macroscopic work adds no collision evaluations.

\subsection{Specific models}\label{sec:models_scheme}
The W iteration, the traction, the form of the particle update and the acceptance are the same for all models, which enter the iteration in three respects. The first is the discrete collision term $C^h(f)$. It has to conserve the quadrature moments of mass, momentum and energy, for otherwise $R_U$ would not be the moment of $R_P$. For asymptotic preservation it must also vanish on the discrete Maxwellian, $C^h(M^h)=0$, for otherwise the equilibrium defect, multiplied by $1/\varepsilon$, would enter at the order of the Euler flux. A discretization lacking the first property is corrected by a Maxwellian-weighted projection, and one lacking the second by subtracting its value at the discrete Maxwellian. The second respect is the velocity space, which is three-dimensional unless the model admits reduced distributions. The third is the rate $1/\widehat\tau$ in (92) and (93). It must bound the fastest relaxation rate of $C^h$, so that the local step is contractive for every $\Delta t_{\rm num}$, and since it multiplies increments only, its value does not affect the converged solution. For the gas models the residuals are $R_U=-\nabla_h\cdot(\bF^h_W+\bF^h_P)$ and (86) with the endpoint $P_e^h=M-W-\tau D^h_W(W)$, and it remains to specify $C^h$, $\widehat\tau$ and the resulting particle update.

\subsubsection{BGK model}
The discrete collision term $C^h(f)=(M-f)/\tau$ is conservative and vanishes on $M$ by the definition of the discrete Maxwellian. The discrete remainder is zero, and $\widehat\tau=\tau$ is the exact rate. The particle residual is
\begin{equation}
 R_P=\frac{P_e^h-P}{\tau}-D^h_P(P)=\frac{M-f}{\tau}-D^h_W(W)-D^h_P(P),
\end{equation}
and by (94) the first particle update reads
\begin{equation}
 \Big[\Big(\frac{1}{\Delta t_{\rm num}}+\frac1\tau\Big)\bI+D^h_{\rm up}\Big]\Delta P=\frac{\widetilde P_e^{h,k+1}-P^k}{\tau}-D^h_P(P^k).
\end{equation}
In the later updates, with $f=\widetilde W^{k+1}+P$ and with $M$ and $\tau$ evaluated from the moments of this $f$,
\begin{equation}
 b^{(n)}=\frac{M-f}{\tau}-D^h_W\big(\widetilde W^{k+1}\big)-D^h_P(P).
\end{equation}
Each particle update is a transport sweep with a known source, and the kinetic cost of one outer iteration is one evaluation of the particle flux and one sweep pair. Where $\eta\gg1$ the iteration is the gas-kinetic Navier--Stokes solver with an exponentially small kinetic correction, and where $\eta\ll1$ it is the discrete-ordinate iteration.

\subsubsection{Shakhov model}
When the flow depends on $d<3$ coordinates, the transverse velocity components $\bxi_t$ are integrated out~[44], and the unknowns are the reduced pair $h=\int f\,\dd\bxi_t$ and $b=\int|\bxi_t|^2f\,\dd\bxi_t$ over the $d$-dimensional velocity space. Both members obey a relaxation equation with the same $\tau$, and the wave, the particle, the endpoint, the residuals and all updates apply to the pair with common coefficients. The discrete collision term is $C^h(f)=(S^h(f)-f)/\tau$, where $S^h$ is the target evaluated with the quadrature heat flux of $f$, and $R^h=(S^h(f)-M)/\tau$. The discrete target is required to conserve the quadrature moments. It satisfies $S^h(M)=M$ whenever the discrete Maxwellian has zero quadrature heat flux, which holds up to the quadrature error of the velocity grid. The linearized model has the relaxation rates $1/\tau$ and $\Pr/\tau$, both bounded by $1/\tau$, so that the rate $\widehat\tau=\tau$ is admissible. The heat-flux correction is kept explicit, and no Jacobian of the target is formed. The particle residual is
\begin{equation}
 R_P=\frac{S^h(f)-M}{\tau}+\frac{P_e^h-P}{\tau}-D^h_P(P)=\frac{S^h(f)-f}{\tau}-D^h_W(W)-D^h_P(P),
\end{equation}
the first particle update reads
\begin{equation}
 \Big[\Big(\frac{1}{\Delta t_{\rm num}}+\frac1\tau\Big)\bI+D^h_{\rm up}\Big]\Delta P=\frac{S^h(f^k)-M^k}{\tau}+\frac{\widetilde P_e^{h,k+1}-P^k}{\tau}-D^h_P(P^k),
\end{equation}
and in the later updates, with $f=\widetilde W^{k+1}+P$ and with the target and $\tau$ evaluated from the moments and the heat flux of this $f$,
\begin{equation}
 b^{(n)}=\frac{S^h(f)-f}{\tau}-D^h_W\big(\widetilde W^{k+1}\big)-D^h_P(P).
\end{equation}
No collision integral is evaluated. With $N_P=m_P=1$, the kinetic cost of one outer iteration is one quadrature of the heat flux, one evaluation of the particle flux and one sweep pair. In the continuum limit, $P=\varepsilon S^{(1),h}+\mathcal O(\varepsilon^2)$.

\subsubsection{ES-BGK model}
The discrete collision term is $C^h(f)=(G^h(f)-f)/\tau$ with $\tau=\mu/(\Pr\,p)$, and $R^h=(G^h(f)-M)/\tau$. The discrete Gaussian $G^h$ is formed from the quadrature moments and the quadrature pressure tensor of $f$ and is required to conserve the quadrature moments. It satisfies $G^h(M)=M$ whenever the quadrature pressure tensor of the discrete Maxwellian is isotropic, which holds up to the quadrature error. Unlike the Shakhov model, this model has relaxation rates that its relaxation time does not bound. The heat flux relaxes at the rate $1/\tau$, but the stress relaxes at the rate $1/(\Pr\,\tau)=p/\mu$, which is the larger of the two for $\Pr<1$. The rate is therefore
\begin{equation}
 \frac{1}{\widehat\tau}=\frac{1}{\Pr\,\tau}=\frac{p}{\mu},
\end{equation}
and the correction of the target is again kept explicit. The particle residual is
\begin{equation}
 R_P=\frac{G^h(f)-M}{\tau}+\frac{P_e^h-P}{\tau}-D^h_P(P)=\frac{G^h(f)-f}{\tau}-D^h_W(W)-D^h_P(P),
\end{equation}
the first particle update reads
\begin{equation}
 \Big[\Big(\frac{1}{\Delta t_{\rm num}}+\frac{1}{\widehat\tau}\Big)\bI+D^h_{\rm up}\Big]\Delta P=R_P^k+\frac{\widetilde P_e^{h,k+1}-P_e^{h,k}}{\widehat\tau},
\end{equation}
and in the later updates $b^{(n)}=(G^h(f)-f)/\tau-D^h_W(\widetilde W^{k+1})-D^h_P(P)$ with $f=\widetilde W^{k+1}+P$. The kinetic cost is that of the Shakhov model, with the quadrature of the pressure tensor in place of that of the heat flux. In the continuum limit, $P=\varepsilon G^{(1),h}+\mathcal O(\varepsilon^2)$. This scheme has not been implemented.

\subsubsection{Boltzmann equation}
The collision term is evaluated by the fast spectral method (FSM)~[45, 46] on a three-dimensional velocity grid. One evaluation costs $\mathcal O(N_{\rm ang}^2N_v\log N_v)$ operations for $N_v$ velocity nodes and $N_{\rm ang}^2$ angular sections. Its output, which includes the factor $1/\varepsilon$, is first corrected by a term $M\,\boldsymbol\gamma\cdot\bpsi$, whose five coefficients $\boldsymbol\gamma$ are fixed by the five conservation conditions, so that the result $\widehat Q^h(f,f)$ satisfies $\sum_j\varpi_j\bpsi_j\widehat Q^h_j=\boldsymbol0$. The collision term used in the residuals is
\begin{equation}
 C^h(f)=Q^h_{\rm FSM}(f,f)=\widehat Q^h(f,f)-\widehat Q^h(M,M),
\end{equation}
which is conservative and has the discrete Maxwellian as an exact equilibrium. The same evaluation provides the loss rate $\Lambda_{i,j}$, the loss part of the collision term divided by $f_{i,j}$. Because the collision term is stiffer than the relaxation term, the rate is
\begin{equation}
 \frac{1}{\widehat\tau_i}=\max\Big(\frac{1}{\tau_i},\ C_{\rm col}\max_j\Lambda_{i,j}\Big),
\end{equation}
with a safety factor $C_{\rm col}\ge1$, evaluated on the state that provides the residual. With $R^h=Q^h_{\rm FSM}(f,f)-(M-f)/\tau$, the particle residual is
\begin{equation}
 R_P=R^h+\frac{P_e^h-P}{\tau}-D^h_P(P)=Q^h_{\rm FSM}(f,f)-D^h_W(W)-D^h_P(P),
\end{equation}
the first particle update reads
\begin{equation}
 \Big[\Big(\frac{1}{\Delta t_{\rm num}}+\frac{1}{\widehat\tau}\Big)\bI+D^h_{\rm up}\Big]\Delta P=R_P^k+\frac{\widetilde P_e^{h,k+1}-P_e^{h,k}}{\widehat\tau},
\end{equation}
followed by one FSM evaluation on $f=\widetilde W^{k+1}+P$, and in the later updates
\begin{equation}
 b^{(n)}=Q^h_{\rm FSM}(f,f)-D^h_W\big(\widetilde W^{k+1}\big)-D^h_P(P).
\end{equation}
The collision operator is thus evaluated $N_P$ times per outer iteration and never in the W iteration. With $N_P=m_P=1$, one outer iteration performs the same kinetic work as one iteration of the conventional iterative scheme (CIS), namely one FSM evaluation, one evaluation of the transport and one sweep pair. In the continuum limit, $P=\varepsilon(f^{(1),h}-W^{(1),h})+\mathcal O(\varepsilon^2)$, where $f^{(1),h}$ and $W^{(1),h}$ are the discrete first-order corrections of the distribution and of the wave.

\subsubsection{Landau equation}
For the Landau equation the scheme is a proposal and has not been implemented. The collision term can be evaluated by the fast spectral method for $Q_L$~[47], at the cost $\mathcal O(N_v\log N_v)$, or by a conservative entropy scheme~[48]. The projection and the equilibrium correction of the Boltzmann case are applied where the scheme requires them, and the result, including the factor $1/\varepsilon$, is denoted $Q^h_L(f,f)$. The W iteration requires only $\mu_L$ and $\kappa_L$. With $R^h=Q^h_L(f,f)-(M-f)/\tau$, the particle residual is
\begin{equation}
 R_P=R^h+\frac{P_e^h-P}{\tau}-D^h_P(P)=Q^h_L(f,f)-D^h_W(W)-D^h_P(P).
\end{equation}
The essential difference from the Boltzmann case lies in the stiffness. Since $Q_L$ is a diffusion operator in velocity, the fastest relaxation rate of its discretization grows like $1/(\varepsilon\,\Delta\xi^2)$ with the velocity mesh size $\Delta\xi$. A scalar rate $1/\widehat\tau$ of this size keeps (93) contractive. The modes that carry the stress and the heat flux, however, relax at the rate $1/\tau$, so that one update with such a rate removes only a fraction $\mathcal O(\Delta\xi^2)$ of their error, and the convergence may be expected to degrade as the velocity grid is refined. In time-dependent schemes the same observation has led to replacing the BGK penalization by a penalization with the Fokker--Planck operator~[49],
\begin{equation}
 \mathcal P_{\rm FP}\,g=\nabla_{\bxi}\cdot\Big(\widetilde M\,\nabla_{\bxi}\frac{g}{\widetilde M}\Big).
\end{equation}
The corresponding choice here is to replace the scalar rate $1/\widehat\tau$ by the operator $\tau^{-1}\bI-(C_{\rm FP}/\varepsilon)\,\mathcal P^h_{\rm FP}$, in the particle update and in the traction alike, where $\widetilde M$ is the Maxwellian of the predicted state and $C_{\rm FP}$ bounds the largest eigenvalue of the diffusion matrix $\int\Phi(\bxi-\bxi_*)f_*\,\dd\bxi_*$. The first particle update then reads
\begin{equation}
 \begin{aligned}
 &\Big[\Big(\frac{1}{\Delta t_{\rm num}}+\frac{1}{\tau}\Big)\bI-\frac{C_{\rm FP}}{\varepsilon}\,\mathcal P^h_{\rm FP}+D^h_{\rm up}\Big]\Delta P=R_P^k+\Theta^k,\\
 &\Theta^k=\Big(\frac{1}{\tau}\bI-\frac{C_{\rm FP}}{\varepsilon}\,\mathcal P^h_{\rm FP}\Big)\big(\widetilde P_e^{h,k+1}-P_e^{h,k}\big),
 \end{aligned}
\end{equation}
followed by one evaluation of $Q^h_L$ on $f=\widetilde W^{k+1}+P$, and in the later updates
\begin{equation}
 b^{(n)}=Q^h_L(f,f)-D^h_W\big(\widetilde W^{k+1}\big)-D^h_P(P).
\end{equation}
The forward and backward sweeps over the cells are retained. In each cell they now require the solution of a linear problem in velocity, which is symmetric positive definite in the unknown $\Delta P/\widetilde M$ and can be solved by conjugate gradients~[49]. Since the operators on the left and in the traction multiply increments only, the fixed point remains the discrete Landau solution, $D^h_W(W)+D^h_P(P)=Q^h_L(f,f)$, independently of $C_{\rm FP}$ and $\widetilde M$.

\subsubsection{Neutron transport}
The scheme for neutron transport follows that of the BGK model and has not been implemented in the present form. The directions are represented by discrete ordinates $\bOm_j$ with weights $\varpi_j$, and $\phi_i=\sum_j\varpi_j\psi_{i,j}$. The discrete collision term is $C^h(\psi)=(S^h-\psi)/\tau$, with the source function (67) formed from the quadrature scalar flux. Its zeroth moment is exactly $c\,(q-\sigma_a\phi)$, the source of the discrete balance law. The macroscopic residual is the residual of this law, and it is again the moment of $R_P$. As for the BGK model, the discrete remainder vanishes and the rate $1/\widehat\tau=1/\tau=c\sigma_t$ is exact. With the endpoint $P_e^h=S-W-\tau D^h_W(W)$, the particle residual is
\begin{equation}
 R_P=\frac{P_e^h-P}{\tau}-D^h_P(P)=\frac{S-\psi}{\tau}-D^h_W(W)-D^h_P(P),
\end{equation}
and the first particle update reads
\begin{equation}
 \Big[\Big(\frac{1}{\Delta t_{\rm num}}+\frac1\tau\Big)\bI+D^h_{\rm up}\Big]\Delta P=\frac{\widetilde P_e^{h,k+1}-P^k}{\tau}-D^h_P(P^k),
\end{equation}
which is a transport sweep with a known source. The W iteration solves the steady balance law for $\phi$ with the particle current frozen,
\begin{equation}\label{eq:W_neutron}
 \begin{aligned}
 &\nabla_h\cdot\bJ^h_W[\phi]+\sigma_a\phi=q-\nabla_h\cdot\bJ^{h,k}_P,\\
 &\bJ^h_W=-\frac{A_1}{3\sigma_t}\nabla_h\frac{\sigma_s\phi+q}{\sigma_t}.
 \end{aligned}
\end{equation}
This is a linear diffusion problem, which can be solved by conjugate gradients or multigrid instead of pseudo-time sweeps. Where $\eta\to0$ the iteration is the source iteration, and where $\eta\to\infty$ the W iteration is the diffusion solve and the particle is exponentially small. The iteration thus has the structure of a source iteration with a diffusion-based acceleration~[3]. It differs from such an acceleration in two respects. Its low-order equation is the balance law itself, closed by the particle current $\bJ_P$, a moment of $P$, rather than a synthetic diffusion correction, and its kinetic unknown is the particle alone. The model problem of the convergence analysis in Section~4 is exactly this equation with $\sigma_a=0$.

\subsubsection{Radiative transfer}
For radiative transfer the particle residual and the particle update are those of neutron transport, with the source function formed from the quadrature scalar flux and the temperature. This scheme, too, has not been implemented in the present form. In a steady state the material equation reduces to $4\pi B(T)=\phi$, the source function becomes $S=\phi/(4\pi)$, and the problem coincides with conservative neutron transport. Absorption followed by re-emission then acts as isotropic scattering with a scattering ratio of one, which is the least favourable case for the source iteration and the one in which acceleration matters most. The temperature follows from $a_RcT^4=\phi$. In a time-dependent computation the iteration is applied within each time step of length $\Delta t_n$, with the time-difference terms added to the residuals and to the diagonals of the two updates, and the W iteration solves the coupled nonlinear system
\begin{equation}
 \left\{
 \begin{aligned}
 &\frac{\phi-\phi^n}{c\,\Delta t_n}+\nabla_h\cdot\bJ^h_W[\phi,T]+\sigma_a\big(\phi-a_RcT^4\big)=-\nabla_h\cdot\bJ^{h,k}_P,\\
 &C_v\,\frac{T-T^n}{\Delta t_n}=\sigma_a\big(\phi-a_RcT^4\big),
 \end{aligned}
 \right.
\end{equation}
for $\phi$ and $T$ with the particle current frozen. In the optically thick regime this system is the discrete equilibrium diffusion equation, and the particle is exponentially small.
\section{Properties}\label{sec:properties}
The first property concerns the limit of the iteration rather than its speed. If the discrete collision term is conservative and the compatibility (84) holds, a state is a fixed point of the outer iteration if and only if it satisfies the discrete kinetic equation (87). This equation contains none of $\Delta t_{\rm num}$, $\widehat\tau$, the approximate Jacobians, $N_U$, $m_U$, $N_P$ and $m_P$, so that the parameters of the iteration change its path but not its limit. The next two subsections concern the behaviour of this limit as $\varepsilon\to0$ and the rate at which it is approached, and precise statements and proofs are given in~[36, 37]. The remaining subsections relate the decomposition and the iteration to the unified gas-kinetic wave--particle method, to the micro--macro decomposition, to penalization and to synthetic acceleration.

\subsection{Asymptotic preservation}\label{sec:ap}
At the continuous level, (27) shows that for every $\eta$ the total flux of the multiscale equations (20) is the Navier--Stokes flux of the model through first order in $\varepsilon$. The same holds at the discrete level under two assumptions, namely that $C^h$ vanishes on discrete Maxwellians and that the horizon is bounded below independently of $\varepsilon$. As $\varepsilon\to0$ on a fixed mesh, the particle then reduces to $P=\varepsilon(f^{(1),h}-W^{(1),h})+\mathcal O(\varepsilon^2)$, and the moments of (87) reduce to the finite-volume Euler equations at leading order and to the gas-kinetic Navier--Stokes scheme with the transport coefficients of the model at first order. The wave flux supplies $\bF_{\rm E}+\bF_{\rm R}$, and the particle supplies $\bF_{\rm S}$. At no point is the mesh required to resolve the mean free path.

\subsection{Acceleration}\label{sec:acceleration}
Two facts explain why the coupled WP iteration converges quickly in the near-continuum regime. The first is that the particle iteration does not undo the macroscopic prediction. Near equilibrium and for every $\eta\ge0$, one particle update with the traction gives $\Delta P=\Delta(e^{-\eta}M)+\mathcal O(\varepsilon)$ and $\Delta W=\Delta[(1-e^{-\eta})M]+\mathcal O(\varepsilon)$, so that $\bU^{k+1}=\widetilde\bU^{k+1}+\mathcal O(\varepsilon)$. The second fact is quantitative. For a linear model problem with wavenumber $\varkappa$ and $z=\varkappa\tau$, one CIS iteration multiplies the error of the density mode by $g_{\rm CIS}=\arctan z/z\ge1-z^2/3$, whereas one outer iteration multiplies it by
\begin{equation}
 g_{\rm WP}\le\frac{z^2}{3}+\frac{\beta(\eta)}{1-\beta(\eta)-3z^2/5},
\end{equation}
where $\beta(\eta)=(1+\eta)e^{-\eta}=1-A_1$. On a fixed mesh $z=\mathcal O(\varepsilon)$ and $\eta\sim\varepsilon^{-1}$. The CIS therefore needs $\mathcal O(\varepsilon^{-2})$ iterations, while $g_{\rm WP}$ is controlled by $\beta$, the share of $\bF_{\rm R}$ carried by the particle in (27), which decays exponentially with $\eta$. For $\eta=4$ and $z=10^{-2}$, for instance, $g_{\rm WP}\le0.10$, whereas $g_{\rm CIS}=1-3.33\times10^{-5}$. In rarefied cells, where $z=\mathcal O(1)$, the CIS itself converges rapidly, and a single macroscopic update per outer iteration suffices.

\subsection{Relation to the UGKWP method}\label{sec:ugkwp}
The unified gas-kinetic wave--particle (UGKWP) method~[24] divides the distribution within each time step $\Delta t$ according to the same integral solution. For a relaxation model with target $g$ and a relaxation time frozen over the step, the distribution at the end of the step is written as the sum of a collided and an uncollided part,
\begin{equation}\label{eq:ugkwp_split}
 \begin{aligned}
 &f(\bx,\bxi,t_{n+1})=f_c+f_u,\\
 &f_u=e^{-\Delta t/\tau}f(\bx-\bxi\Delta t,\bxi,t_n),\\
 &f_c\approx c_1\,g+c_2\big(\partial_tg+\bxi\cdot\nabla_{\bx}g\big),
 \end{aligned}
\end{equation}
with $c_1=1-e^{-\Delta t/\tau}$ and $c_2=\Delta t\,e^{-\Delta t/\tau}-\tau(1-e^{-\Delta t/\tau})$. The collided part is the equilibrium integral over the step, with the target expanded linearly along the characteristic about the current point. It is carried by its moments, as the hydrodynamic part of the cell, while the uncollided part is carried by stochastic particles. Comparison with (9) and (82) shows that this split is the decomposition with the horizon equal to the time step. For $\ell=\Delta t$ and $\eta=\Delta t/\tau$ the particle $e^{-\nu(\ell)}f_\ell$ is $f_u$, and since $A_0(\eta)=c_1$ and $A_1(\eta)\,\tau=-c_2$, the discrete wave (82) is $f_c$. The correspondence extends to the evolution. Integrating the particle equation (18) with $R=0$ and $\ell\equiv\Delta t$ along a characteristic over one step gives
\begin{equation}\label{eq:ugkwp_step}
 P(\bx,\bxi,t_{n+1})=e^{-\Delta t/\tau}\big[P+W\big](\bx-\bxi\Delta t,\bxi,t_n),
\end{equation}
in which the transport and the attenuation $-P/\tau$ produce the first term, the free flight of the particles that survive the step, and the exchange $\cS_\ell$ integrated over the step produces the second, the collisionless fraction $e^{-\Delta t/\tau}$ of the wave. The wave at $t_{n+1}$ is, by (9), the equilibrium integral over the step alone, that is, the molecules that collided during the step. These are the three operations of the UGKWP method, in which the surviving particles are followed, the colliding particles are absorbed into the wave, and the collisionless fraction of the wave is re-sampled as particles. The UGKWP method may therefore be read as a time discretization of the multiscale equations (20) with the horizon equal to the time step, in which the particle equation is solved by Monte Carlo particles and the wave is carried by its moments.

The two approaches differ first in the horizon. In the UGKWP method the split is made in the numerical update, and its proportion is set by the global time step, so that the smallest cell of a mesh fixes the particle fraction everywhere. Here the horizon is a field $\ell(\bx,t)$, the decomposition is made in the equation, and the two components obey the exact equations (13) and (18), in which the motion of the horizon enters through the factor $1-\partial_t\ell-\bxi\cdot\nabla_{\bx}\ell$ of the exchange. They differ, secondly, in the collision term. The wave of the UGKWP method is the equilibrium integral of the relaxation target of the model, and the method is formulated for relaxation models. Here the wave integrates the Maxwellian, and the identity (4) carries the same wave to every collision operator through the remainder $R$, which is confined to the particle. The third difference lies in the representation. The UGKWP method follows stochastic particles, samples the collisionless fraction of the wave from the Maxwellian of the hydrodynamic part, defines the wave as the remainder of the particles, and computes the wave flux from the integral solution of the unified gas-kinetic scheme, including the free transport of the wave within the step. Here the wave is a functional of $\bU$ and the particle is the remainder. The particle is the discrete-ordinate solution of its own equation, the exchange carries the actual shape of the wave, and the wave flux is the gas-kinetic Navier--Stokes flux weighted by $A_0$ and $A_1$, a different discretization of the same flux $\bF_W$ that agrees with it through first order in $\tau$. Finally, the UGKWP method advances the solution in time, whereas the coupled WP iteration of Section~3 solves the steady equations, with the fixed point and the convergence factor discussed above. The two methods agree in the limit of mesh refinement and, with $\ell=\Delta t$, carry the same particle fraction. The effect of replacing the global time step by a local horizon is illustrated in Section~5.

\subsection{Relation to the micro--macro decomposition}\label{sec:micromacro}
The micro--macro decomposition~[9, 10, 11] writes $f=M+g$, with the Maxwellian $M=M[\bU]$ of the conservative variables and a kinetic remainder $g$ that has vanishing conservative moments, $\avg{\bpsi g}=\boldsymbol0$, and is obtained by projecting the kinetic equation onto the complement of the collision invariants. The conservation law is closed by the Euler flux of $M$ and by the flux of $g$, which carries the whole non-equilibrium transport, and the equation of $g$ retains the collision operator in full. Since also $f=W+P$, the two decompositions are related exactly by
\begin{equation}\label{eq:mm_relation}
 P=g+(M-W),
\end{equation}
and by (23) the difference is $M-W=e^{-\eta}M+A_1\tau\,\cD M+\mathcal O(\tau^2)$. Three consequences follow. The particle carries conservative moments, $\avg{\bpsi P}=e^{-\eta}\bU+\mathcal O(\varepsilon^2)$, whereas $g$ carries none. For $\eta=0$ the particle is $f$ itself. For $\eta\to\infty$ the particle is $P=g+\varepsilon\bar\tau\cD^{(0)}M+\mathcal O(\varepsilon^2)=\varepsilon r^{(1)}+\mathcal O(\varepsilon^2)$, so that of the kinetic remainder $g=\varepsilon f^{(1)}+\mathcal O(\varepsilon^2)$ it retains only the deviation of the model from the relaxation response, and it vanishes for the BGK model. Accordingly, the flux carried by the kinetic unknown in the conservation law is $\varepsilon\bF^{(1)}$, the entire first-order flux, in the micro--macro decomposition, but only $(1+\eta)e^{-\eta}\bF_{\rm R}+\bF_{\rm S}$ in (20), the relaxation part $A_1\bF_{\rm R}$ having been moved into the wave flux, which is a functional of $\bU$. This transfer of the relaxation transport from the kinetic to the macroscopic unknown is the source of the acceleration described in Section~4.2, and it is what allows the macroscopic iteration to act as a Navier--Stokes solver. It comes at a price. The particle is not orthogonal to the invariants, so that the consistency $\bU=\avg{\bpsi(W+P)}$ has to be maintained. At the continuous level this is guaranteed by the equivalence stated in Section~2.5, and at the discrete level by the acceptance (95).

\subsection{Relation to penalization}\label{sec:penalization}
Filbet and Jin~[8] remove the stiffness of a general collision operator by adding and subtracting a BGK operator with a rate $\lambda$, $\varepsilon^{-1}Q(f)=\varepsilon^{-1}[Q(f)-\lambda(M-f)]+\varepsilon^{-1}\lambda(M-f)$, and by treating the penalty $\varepsilon^{-1}\lambda(M-f)$ implicitly and the penalized operator explicitly in time. With $1/\tau=\lambda/\varepsilon$, the penalized operator is the collision remainder $R$ of (4), and the identity underlying the decomposition is the penalization identity. The two constructions use it differently. In the penalization the unknown remains $f$, the identity distributes the collision term between an implicit and an explicit time integration, the rate $\lambda$ must dominate the stiffness of $Q$ for the explicit part to be stable, and the asymptotic-preserving property follows from the implicit penalty. In the decomposition the identity defines the unknowns. The relaxation term is integrated along the characteristics into the wave, a functional of $\bU$, and only the remainder is carried by the particle. The relaxation time is the physical one, so that with $\tau=\mu/p$ the relaxation flux $\bF_{\rm R}$ is the physical viscous flux and the wave flux is a Navier--Stokes flux of the model. The stiffness of the remainder enters only through the preconditioning rate $1/\widehat\tau$ of the P iteration, which does not affect the converged solution. In the steady iteration, the implicit operator $(1/\Delta t_{\rm num}+1/\widehat\tau)\bI+D^h_{\rm up}$ of (93) is the counterpart of the implicit penalty, and the Fokker--Planck penalization of~[49] corresponds to the operator-valued rate proposed for the Landau equation in Section~3.6.

\subsection{Relation to synthetic acceleration}\label{sec:synthetic}
Synthetic acceleration corrects each transport sweep of the source iteration by the solution of a low-order equation. In diffusion synthetic acceleration~[28, 29, 3] the low-order equation is a diffusion equation for the error of the scalar flux, driven by the residual of the sweep, and the spectral radius of the iteration is bounded independently of the mesh in the diffusive limit. In the general synthetic iterative scheme~[33, 34] the low-order equations are the moment equations of the kinetic equation, closed by the stress and heat flux of the current kinetic iterate through their deviation from the Navier--Stokes laws, and the updated conservative variables define the equilibrium of the next kinetic iteration. In the implicit UGKS~[30, 32], implicit macroscopic equations with fluxes evaluated from the current distribution predict the conservative variables before the implicit update of the distribution. The coupled WP iteration shares this macroscopic--microscopic structure but differs in three respects. Its low-order equation is the total conservation law itself, closed by the wave flux, a functional of $\bU$ that contains the fraction $A_1$ of the relaxation transport, and by the lagged particle flux. What is lagged from the kinetic iterate is therefore not the full deviation from the Navier--Stokes laws but the particle share $(1+\eta)e^{-\eta}\bF_{\rm R}+\bF_{\rm S}$, which decays with the horizon-to-relaxation ratio, and Section~4.2 quantifies the resulting convergence factor. Its kinetic unknown is the particle alone rather than the distribution function. And the macroscopic prediction reaches the kinetic equation through the endpoint traction and the fixed predicted wave, rather than through the equilibrium of the next sweep. For linear transport with $\eta\to\infty$, the W iteration (116) is a diffusion solve for the scalar flux itself, with Fick's current weighted by $A_1$, and the particle is exponentially small. At finite $\eta$ the low-order equation contains the particle current, so that the iteration passes continuously from the source iteration at $\eta\to0$ to the diffusion solve.

\section{Numerical illustration}\label{sec:illustration}
Three configurations illustrate the framework. Hypersonic flow past a circular cylinder contains, within one domain, a bow shock, a stagnation region, a boundary layer on a diffuse wall and a rarefied wake, and it is computed here from the transitional to the continuum regime. Hypersonic flow around the X38 vehicle carries these features over to a three-dimensional geometry. The lid-driven cavity at the Reynolds number $\mathrm{Re}=1000$ is a continuum flow in which the conventional iteration is at its slowest. We first show flow fields, then the kinetic degrees of freedom that the decomposition requires, and finally the convergence of the steady iteration. The computations shown in this section are taken from~[35] and~[37], where the schemes, the settings and the full assessment are documented. The discussion here is confined to the features that follow from the structure of Sections~2--4.

We begin with a three-dimensional configuration. Fig.~3 shows the temperature around the X38 vehicle for the Shakhov model of a monatomic gas, at a free-stream Mach number of $5$ and an angle of attack of $15^\circ$, with diffuse reflection at the vehicle surface and on an unstructured polyhedral mesh of $560593$ cells. At the Knudsen number $\mathrm{Kn}=10^{-1}$ the heated region is broad and diffuse and encloses the whole vehicle, with the highest temperatures on the windward side below the nose. At $\mathrm{Kn}=10^{-4}$ the shock layer is thin, the temperature peaks sharply at the nose, and a narrow heated wake trails the base of the vehicle. Both computations use the same mesh and the same algorithm, and the passage between the two regimes is carried entirely by the local ratio $\eta$, which assigns a growing share of the distribution to the wave as the Knudsen number decreases. The example serves as a qualitative demonstration of the framework on a complex geometry rather than as a quantitative aerodynamic validation.
\begin{figure}[htbp]
\centering
\includegraphics[width=\textwidth]{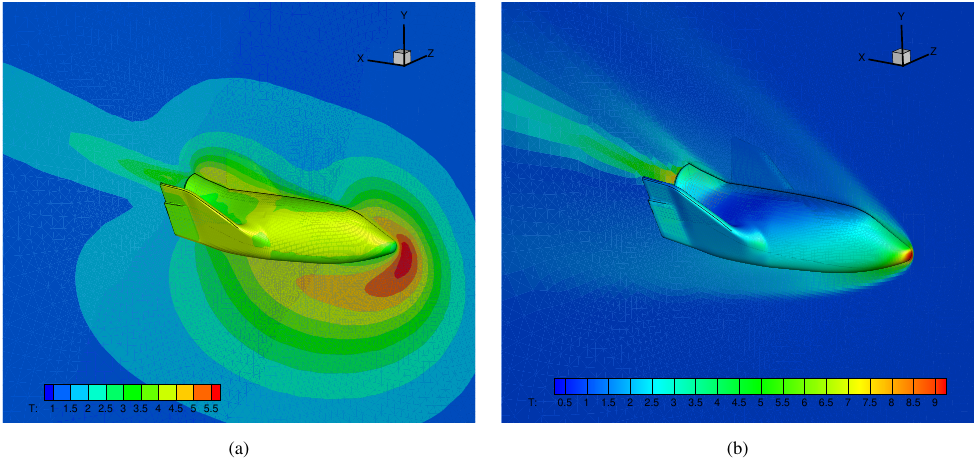}
\caption{Temperature around the X38 vehicle at Mach 5 and an angle of attack of $15^\circ$. (a) $\mathrm{Kn}=10^{-1}$. (b) $\mathrm{Kn}=10^{-4}$. The temperature is shown on the vehicle surface and in the surrounding flow field for the Shakhov model, and both computations use the same algorithm on the same mesh.}\label{fig:x38}
\end{figure}

The Boltzmann collision operator enters the same equations through its remainder alone. Fig.~4 shows the temperature of Mach-5 flow past a cylinder of unit radius obtained by the coupled WP iteration of Section~3 with the Boltzmann collision operator, evaluated by the fast spectral method on a three-dimensional velocity grid, at the Knudsen numbers $\varepsilon=1$, $10^{-2}$ and $10^{-4}$. The body-fitted mesh has $129\times80$ cells at $\varepsilon=1$ and $129\times40$ cells at the two smaller Knudsen numbers, and the velocity grid has $96\times96\times24$ nodes at the two larger Knudsen numbers and $32\times32\times12$ nodes at $\varepsilon=10^{-4}$. The reference, drawn as lines at the same levels, is the solution of the conventional iterative scheme (CIS) at all three Knudsen numbers, and the relative $L_2$ differences printed above the panels are $1.076\times10^{-5}$, $7.867\times10^{-4}$ and $1.275\times10^{-3}$. The three computations use the same equations and the same iteration. Only the controls of Section~3 change, from one macroscopic update per outer iteration at the two larger Knudsen numbers to $120$ at the smallest, where the W iteration does the work of a Navier--Stokes solver and the particle supplies the correction flux $\bF_{\rm S}$ of (28).
\begin{figure}[htbp]
\centering
\includegraphics[width=\textwidth]{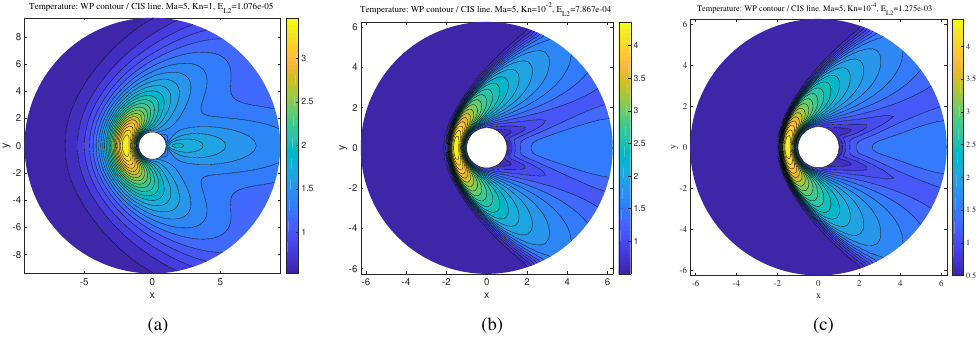}
\caption{Temperature of Mach-5 flow past a cylinder with the Boltzmann collision operator. (a) $\varepsilon=1$. (b) $\varepsilon=10^{-2}$. (c) $\varepsilon=10^{-4}$. Filled contours show the coupled WP iteration and lines at the same levels the conventional iterative scheme. The relative $L_2$ difference is printed above each panel.}\label{fig:temperature}
\end{figure}

Behind such flow fields lies the division of the distribution between wave and particle, which decides how many kinetic degrees of freedom a computation needs. Fig.~5 shows the equivalent particle number per cell, the mass carried by the particle divided by a fixed reference particle weight, in a Monte Carlo realization of the decomposition for the Shakhov relaxation model. The flow is Mach-8 flow past a cylinder of unit radius at $\mathrm{Kn}=10^{-3}$, computed on a body-fitted mesh of $100\times64$ cells that extends to fifteen radii. The horizon is the cell-crossing time, $C_\ell=1$ in (81), and the field is compared with that of the UGKWP method~[24] on the same mesh, in which the particle fraction $e^{-\Delta t/\tau}$ is set by the global time step. With the local horizon, the particles are confined to the shock layer and the near wake, where $\eta$ is of order one, and they have disappeared from the rest of the domain. With the global time step, the far field, where the cells are large, retains a substantial particle population. At $\mathrm{Kn}=1$ the two fields coincide, since the particle fraction is close to one throughout. As the Knudsen number decreases, the local horizon reduces the effective kinetic degrees of freedom relative to the UGKWP method, by factors of about $2.6$ at $\mathrm{Kn}=10^{-2}$ and about $88$ at $\mathrm{Kn}=10^{-3}$. This is the adaptivity of the continuous spectrum of Section~2.6. The same equations are solved everywhere, and the local ratio $\eta$ decides how much of the distribution is carried by the wave.
\begin{figure}[htbp]
\centering
\includegraphics[width=0.92\textwidth]{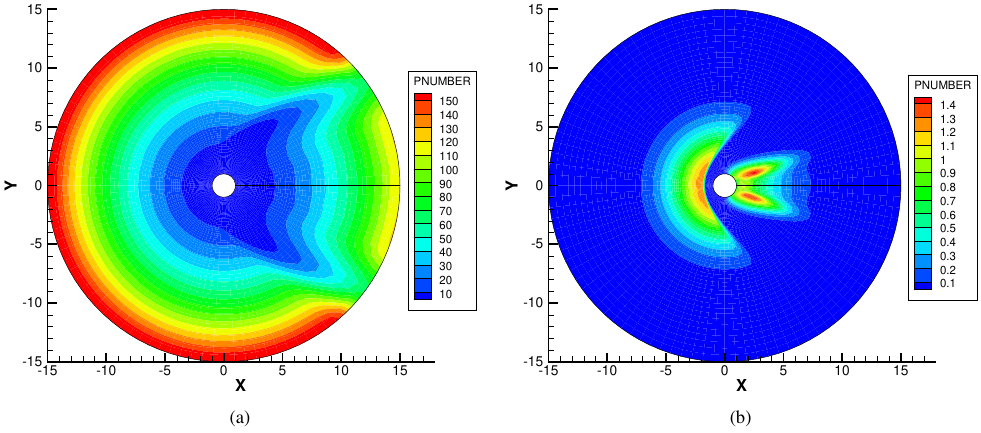}
\caption{Equivalent particle number per cell in Mach-8 flow past a cylinder at $\mathrm{Kn}=10^{-3}$. (a) Global time step of the UGKWP method. (b) Local horizon of the wave--particle decomposition. Both computations use the Shakhov model, and the colour scales of the two panels differ.}\label{fig:pnumber}
\end{figure}

Finally, we turn to the acceleration. Fig.~6 shows the histories of the normalized macroscopic residual against the outer iteration for the lid-driven cavity at $\mathrm{Re}=1000$ and for the cylinder at $\varepsilon=10^{-4}$. The coupled WP iteration, labelled WP in the figure, is compared with the CIS, applied here to the Shakhov model, and with the Navier--Stokes solver, for which the abscissa counts implicit steps. The ratios of the numbers of iterations to convergence, CIS to WP, printed in the panels are $3797.50$ and $54412.11$. This is the acceleration anticipated in Section~4.2. The conventional iteration needs $\mathcal O(\varepsilon^{-2})$ iterations, whereas the convergence factor of the coupled WP iteration is controlled by the particle share $(1+\eta)e^{-\eta}$ of the transport. Because each outer iteration contains $N_U$ macroscopic updates, the number of outer iterations alone does not measure the work relative to the Navier--Stokes solver.
\begin{figure}[htbp]
\centering
\includegraphics[width=\textwidth]{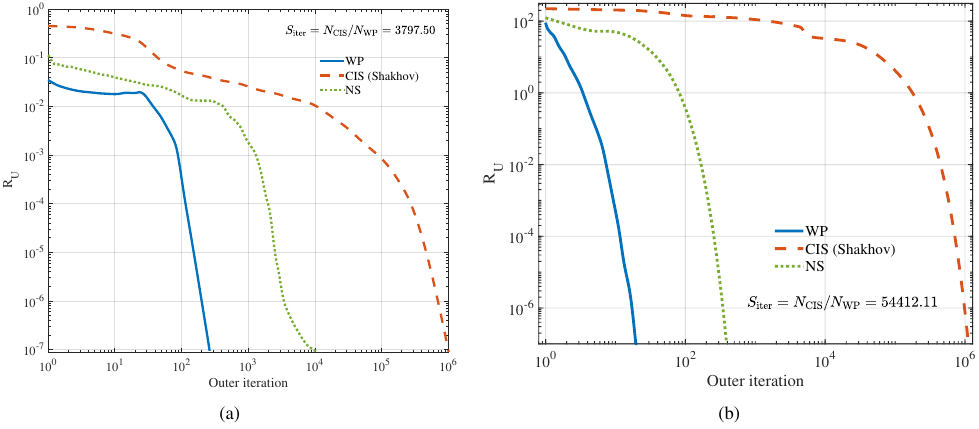}
\caption{Histories of the normalized macroscopic residual against the outer iteration. (a) Lid-driven cavity at $\mathrm{Re}=1000$. (b) Cylinder at $\varepsilon=10^{-4}$. The curves show the coupled WP iteration (WP), the conventional iterative scheme of the Shakhov model (CIS) and the Navier--Stokes solver (NS), whose abscissa counts its implicit steps. The ratio of the numbers of iterations to convergence, CIS to WP, is printed in each panel.}\label{fig:residual}
\end{figure}

\FloatBarrier
\section{Conclusions}\label{sec:summary}
The wave--particle decomposition writes the distribution function as $f=W+P$, where the wave (9) is a functional of the conservative variables and the particle is its exact complement. The two components satisfy the kinetic equations (13) and (18), whose moments are the extended Navier--Stokes equations (17) and (19). The conservation laws and the particle equation form the closed system (20), which is equivalent to the kinetic equation for every horizon and divides the fluxes between wave and particle according to (27). The ratio $\eta$ thereby parametrizes a continuous spectrum from the kinetic equation to the Navier--Stokes equations. The coupled WP iteration alternates the W iteration for $\bU$ and the P iteration for $P$, linked by the endpoint traction and the acceptance. Its fixed point is the discrete kinetic solution, it is asymptotic preserving, and its near-continuum convergence factor is bounded by $(1+\eta)e^{-\eta}$.
\begin{table}[tbp]
\centering\footnotesize
\setlength{\tabcolsep}{2.5pt}
\caption{Model-dependent parts}\label{tab:summary}
\begin{tabular}{@{}llllll@{}}
\toprule
Model & $\tau$ & Remainder $R$ & Particle, $\eta\to\infty$ & Evaluation of $C^h$ & Rate or operator\\
\midrule
BGK & $\mu/p$ & $0$ & none & none & $1/\tau$\\
Shakhov & $\mu/p$ & $(S-M)/\tau$ & $(1-\Pr)\,\qNS$ & quadrature of $\bq$ & $1/\tau$\\
ES-BGK & $\mu/(\Pr\,p)$ & $(G-M)/\tau$ & $(1-1/\Pr)\,\sNS$ & quadrature of $\boldsymbol\Theta$ & $1/(\Pr\,\tau)$\\
Boltzmann & $\mu/p$ & $Q/\varepsilon-(M-f)/\tau$ & $(1-\Pr)\,\qNS$ & FSM & $\max(1/\tau,\,C_{\rm col}\max_j\Lambda_j)$\\
Landau & $\mu_L/p$ & $Q_L/\varepsilon-(M-f)/\tau$ & $(1-\Pr)\,\qNS$ & spectral or entropy & $1/\tau-(C_{\rm FP}/\varepsilon)\,\mathcal P_{\rm FP}$\\
Neutrons & $1/(c\sigma_t)$ & $0$ & none & quadrature of $\phi$ & $c\sigma_t$\\
Radiation & $1/(c\sigma_t)$ & $0$ & none & quadrature of $\phi$ & $c\sigma_t$\\
\bottomrule
\end{tabular}
\tablelegend{The fourth column is the flux carried by the particle as $\eta\to\infty$, and the last column is the rate, or operator, used in the traction and in the particle update. In the last two rows the source function $S$ replaces the Maxwellian.}
\end{table}

The collision model enters the theory through the remainder, and the algorithm through the discrete collision term, the velocity space and one stiffness rate. Table~2 collects these parts for the seven models. For the relaxation models the remainder is explicit or absent, and the particle supplies the Prandtl-number correction, of the heat flux for the Shakhov model and of the stress for the ES-BGK model. For the Boltzmann and Landau equations the remainder contains the full collision operator, and the relaxation term serves only to define the wave. For neutron transport and radiative transfer the source function replaces the Maxwellian and the diffusion equation replaces the Navier--Stokes equations. A further model requires its remainder, a discrete collision term with the two properties stated in Section~3.6, a velocity representation and a bound of its fastest relaxation rate. In this sense the wave--particle decomposition is a framework for multiscale transport, a single set of continuous-spectrum equations and a single coupled WP iteration into which a transport model enters through its remainder, its discrete collision term and one rate.
The computations of Section~5 show what this structure means in practice for aerodynamic flows. The kinetic representation is confined to the regions in which the horizon is short compared with the relaxation time, the same algorithm carries over to three-dimensional vehicles, and in the near-continuum regime the steady iteration converges at a rate that the conventional iteration cannot approach.
\FloatBarrier
\section*{Abbreviations}
\noindent
\begin{tabular}{@{}ll@{}}
BGK & Bhatnagar--Gross--Krook\\
CIS & Conventional iterative scheme\\
ES-BGK & Ellipsoidal statistical BGK\\
FSM & Fast spectral method\\
LU--SGS & Lower--upper symmetric Gauss--Seidel\\
NS & Navier--Stokes\\
PR & Point relaxation\\
UGKS & Unified gas-kinetic scheme\\
UGKWP & Unified gas-kinetic wave--particle\\
WP & Wave--particle
\end{tabular}

\section*{Declarations}
\subsection*{Availability of data and materials}
The datasets used and analysed during the current study are available from the corresponding author on reasonable request.

\subsection*{Competing interests}
KX is one of the Editors-in-Chief of Advances in Aerodynamics and takes no part in the editorial handling of this manuscript. The authors declare that they have no other competing interests.

\subsection*{Funding}
Chang Liu is partially supported by the Beijing Natural Science Foundation (Z230003), the National Natural Science Foundation of China (12031001, 12102061), and the Presidential Foundation of the China Academy of Engineering Physics (YZJJZQ2022017). The authors are partially supported by the National Key R\&D Program of China (2022YFA1004500). Kun Xu is partially supported by the National Natural Science Foundation of China (12172316, 92371107) and the Hong Kong Research Grants Council (16301222, 16208324).

\subsection*{Authors' contributions}
CL developed the theory and the coupled WP iteration, carried out the computations and drafted the manuscript. KX conceived the wave--particle methodology, supervised the work and revised the manuscript. Both authors read and approved the final manuscript.

\subsection*{Use of large language models}
A large language model (Claude, Anthropic) was used for language polishing only. The core content of the manuscript, including the formulation, derivations, algorithms and results, was written by the authors, who reviewed the polished text and take full responsibility for it.

\subsection*{Acknowledgements}
Not applicable.


\begin{thebibliography}{99}
\bibitem{jin1999} Jin S (1999) Efficient asymptotic-preserving (AP) schemes for some multiscale kinetic equations. SIAM J Sci Comput 21:441--454
\bibitem{jin2012} Jin S (2012) Asymptotic preserving (AP) schemes for multiscale kinetic and hyperbolic equations: a review. Riv Mat Univ Parma 3:177--216
\bibitem{adams2002} Adams ML, Larsen EW (2002) Fast iterative methods for discrete-ordinates particle transport calculations. Prog Nucl Energy 40:3--159
\bibitem{jin2000} Jin S, Pareschi L, Toscani G (2000) Uniformly accurate diffusive relaxation schemes for multiscale transport equations. SIAM J Numer Anal 38:913--936
\bibitem{klar1998} Klar A (1998) An asymptotic-induced scheme for nonstationary transport equations in the diffusive limit. SIAM J Numer Anal 35:1073--1094
\bibitem{pareschi2005} Pareschi L, Russo G (2005) Implicit--explicit Runge--Kutta schemes and applications to hyperbolic systems with relaxation. J Sci Comput 25:129--155
\bibitem{dimarco2014} Dimarco G, Pareschi L (2014) Numerical methods for kinetic equations. Acta Numer 23:369--520
\bibitem{filbet2010} Filbet F, Jin S (2010) A class of asymptotic-preserving schemes for kinetic equations and related problems with stiff sources. J Comput Phys 229:7625--7648
\bibitem{liu2004} Liu T-P, Yu S-H (2004) Boltzmann equation: micro-macro decompositions and positivity of shock profiles. Commun Math Phys 246:133--179
\bibitem{bennoune2008} Bennoune M, Lemou M, Mieussens L (2008) Uniformly stable numerical schemes for the Boltzmann equation preserving the compressible Navier--Stokes asymptotics. J Comput Phys 227:3781--3803
\bibitem{lemou2008} Lemou M, Mieussens L (2008) A new asymptotic preserving scheme based on micro-macro formulation for linear kinetic equations in the diffusion limit. SIAM J Sci Comput 31:334--368
\bibitem{gamba2019} Gamba IM, Jin S, Liu L (2019) Micro-macro decomposition based asymptotic-preserving numerical schemes and numerical moments conservation for collisional nonlinear kinetic equations. J Comput Phys 382:264--290
\bibitem{degond2005} Degond P, Jin S, Mieussens L (2005) A smooth transition model between kinetic and hydrodynamic equations. J Comput Phys 209:665--694
\bibitem{degond2010} Degond P, Dimarco G, Mieussens L (2010) A multiscale kinetic--fluid solver with dynamic localization of kinetic effects. J Comput Phys 229:4907--4933
\bibitem{xu2010} Xu K, Huang J-C (2010) A unified gas-kinetic scheme for continuum and rarefied flows. J Comput Phys 229:7747--7764
\bibitem{huang2012ugks} Huang J-C, Xu K, Yu P (2012) A unified gas-kinetic scheme for continuum and rarefied flows II: multi-dimensional cases. Commun Comput Phys 12:662--690
\bibitem{xu2014direct} Xu K (2015) Direct modeling for computational fluid dynamics: construction and application of unified gas-kinetic schemes. World Scientific, Singapore
\bibitem{guo2013} Guo Z, Xu K, Wang R (2013) Discrete unified gas kinetic scheme for all Knudsen number flows: low-speed isothermal case. Phys Rev E 88:033305
\bibitem{mieussens2013} Mieussens L (2013) On the asymptotic preserving property of the unified gas kinetic scheme for the diffusion limit of linear kinetic models. J Comput Phys 253:138--156
\bibitem{liu2016boltzmann} Liu C, Xu K, Sun Q, Cai Q (2016) A unified gas-kinetic scheme for continuum and rarefied flows IV: full Boltzmann and model equations. J Comput Phys 314:305--340
\bibitem{sun2015} Sun W, Jiang S, Xu K (2015) An asymptotic preserving unified gas kinetic scheme for gray radiative transfer equations. J Comput Phys 285:265--279
\bibitem{liu2017plasma} Liu C, Xu K (2017) A unified gas kinetic scheme for continuum and rarefied flows V: multiscale and multi-component plasma transport. Commun Comput Phys 22:1175--1223
\bibitem{liu2019multiphase} Liu C, Wang Z, Xu K (2019) A unified gas-kinetic scheme for continuum and rarefied flows VI: dilute disperse gas-particle multiphase system. J Comput Phys 386:264--295
\bibitem{liu2020ugkwp} Liu C, Zhu Y, Xu K (2020) Unified gas-kinetic wave-particle methods I: continuum and rarefied gas flow. J Comput Phys 401:108977
\bibitem{zhu2019ugkwp} Zhu Y, Liu C, Zhong C, Xu K (2019) Unified gas-kinetic wave-particle methods II: multiscale simulation on unstructured mesh. Phys Fluids 31:067105
\bibitem{li2020photon} Li W, Liu C, Zhu Y, Zhang J, Xu K (2020) Unified gas-kinetic wave-particle methods III: multiscale photon transport. J Comput Phys 408:109280
\bibitem{guo2026representation} Guo Z, Zhu Y, Xu K (2026) Kinetic representation of the unified gas-kinetic wave-particle method and beyond. Commun Comput Phys 39:1512--1535
\bibitem{alcouffe1977} Alcouffe RE (1977) Diffusion synthetic acceleration methods for the diamond-differenced discrete-ordinates equations. Nucl Sci Eng 64:344--355
\bibitem{larsen1982} Larsen EW (1982) Unconditionally stable diffusion-synthetic acceleration methods for the slab geometry discrete ordinates equations. Part I: theory. Nucl Sci Eng 82:47--63
\bibitem{zhu2016implicit} Zhu Y, Zhong C, Xu K (2016) Implicit unified gas-kinetic scheme for steady state solutions in all flow regimes. J Comput Phys 315:16--38
\bibitem{zhu2017multigrid} Zhu Y, Zhong C, Xu K (2017) Unified gas-kinetic scheme with multigrid convergence for rarefied flow study. Phys Fluids 29:096102
\bibitem{xu2022implicit} Xu X, Zhu Y, Liu C, Xu K (2022) UGKS-based implicit iterative method for multiscale nonequilibrium flow simulations. SIAM J Sci Comput 44:B996--B1017
\bibitem{su2020} Su W, Zhu L, Wang P, Zhang Y, Wu L (2020) Can we find steady-state solutions to multiscale rarefied gas flows within dozens of iterations? J Comput Phys 407:109245
\bibitem{zhu2021gsis} Zhu L, Pi X, Su W, Li Z-H, Zhang Y, Wu L (2021) General synthetic iterative scheme for nonlinear gas kinetic simulation of multi-scale rarefied gas flows. J Comput Phys 430:110091
\bibitem{liu_wpd_I} Liu C, Xu K (2026) Wave--particle decomposition for kinetic equations I: theory and numerics. Preprint, arXiv:2606.26915
\bibitem{liu_shakhov} Liu C, Xu K (2026) An implicit scheme for the wave--particle decomposition of the Shakhov kinetic model. Preprint
\bibitem{liu_boltzmann} Liu C, Xu K (2026) Wave--particle decomposition for kinetic equations II: full Boltzmann equation. Preprint
\bibitem{bgk1954} Bhatnagar PL, Gross EP, Krook M (1954) A model for collision processes in gases. I. Small amplitude processes in charged and neutral one-component systems. Phys Rev 94:511--525
\bibitem{shakhov1968} Shakhov EM (1968) Generalization of the Krook kinetic relaxation equation. Fluid Dyn 3:95--96
\bibitem{holway1966} Holway LH (1966) New statistical models for kinetic theory: methods of construction. Phys Fluids 9:1658--1673
\bibitem{liu2025neutron} Liu G, Tan S, Wang Y (2025) The unified gas kinetic wave-particle method for the neutron transport equation. Preprint, arXiv:2509.10178
\bibitem{xu2001} Xu K (2001) A gas-kinetic BGK scheme for the Navier--Stokes equations and its connection with artificial dissipation and Godunov method. J Comput Phys 171:289--335
\bibitem{yoon1988} Yoon S, Jameson A (1988) Lower-upper symmetric-Gauss-Seidel method for the Euler and Navier--Stokes equations. AIAA J 26:1025--1026
\bibitem{chu1965} Chu CK (1965) Kinetic-theoretic description of the formation of a shock wave. Phys Fluids 8:12--22
\bibitem{mouhot2006} Mouhot C, Pareschi L (2006) Fast algorithms for computing the Boltzmann collision operator. Math Comput 75:1833--1852
\bibitem{wu2013} Wu L, White C, Scanlon TJ, Reese JM, Zhang Y (2013) Deterministic numerical solutions of the Boltzmann equation using the fast spectral method. J Comput Phys 250:27--52
\bibitem{pareschi2000} Pareschi L, Russo G, Toscani G (2000) Fast spectral methods for the Fokker--Planck--Landau collision operator. J Comput Phys 165:216--236
\bibitem{degond1994} Degond P, Lucquin-Desreux B (1994) An entropy scheme for the Fokker--Planck collision operator of plasma kinetic theory. Numer Math 68:239--262
\bibitem{jin2011} Jin S, Yan B (2011) A class of asymptotic-preserving schemes for the Fokker--Planck--Landau equation. J Comput Phys 230:6420--6437
\end{thebibliography}
\end{document}